\documentclass[fleqn,usenatbib]{mnras}
\usepackage{newtxtext,newtxmath}
\usepackage[T1]{fontenc}
\DeclareRobustCommand{\VAN}[3]{#2}
\let\VANthebibliography\thebibliography
\def\thebibliography{\DeclareRobustCommand{\VAN}[3]{##3}\VANthebibliography}

\DeclareRobustCommand{\DE}[3]{#2}
\let\DEthebibliography\thebibliography
\def\thebibliography{\DeclareRobustCommand{\DE}[3]{##3}\DEthebibliography}

\usepackage{graphicx}	
\usepackage{amsmath}	
\usepackage[normalem]{ulem} 

\newcommand{\Msun}{$\mathrm{M}_{\odot}$}

\title[Large-scale driving on the ISM]{
The signature of large scale turbulence driving on the structure of the interstellar medium}

\author[T. Colman et al.]{
Tine Colman,$^{1}$\thanks{E-mail: tine.colman@cea.fr}
Jean-Fran\c{c}ois Robitaille,$^{2}$
Patrick Hennebelle,$^{1}$
Marc-Antoine Miville-Desch\^enes,$^{1}$
\newauthor No\'e Brucy,$^{1}$ 
Ralf~S.~Klessen,$^{3,4}$
Simon~C.~O.~Glover,$^{3}$
Juan~D.~Soler,$^{5}$
Davide Elia,$^{5}$
Alessio Traficante,$^{5}$
\newauthor Sergio Molinari,$^{5}$
Leonardo Testi$^{6,7}$
\\
$^{1}$Universit\'e Paris Saclay and Universit\'e de Paris, CEA, CNRS, AIM, F-91190 Gif-sur-Yvette, France\\
$^{2}$Univ. Grenoble Alpes, CNRS, IPAG, 38000 Grenoble, France\\
$^{3}$Universit\"at Heidelberg, Zentrum f\"ur Astronomie, Institut f\"ur Theoretische Astrophysik, Albert-Ueberle-Str 2, D-69120 Heidelberg, Germany\\
$^{4}$Universit\"at Heidelberg, Interdisziplin\"ares Zentrum f\"ur Wissenschaftliches Rechnen, Im Neuenheimer Feld 205, D-69120 Heidelberg, Germany\\
$^{5}$INAF - Istituto di Astrofisica e Planetologia Spaziali, via Fosso del Cavaliere 100, 00133, Roma, Italy \\
$^{6}$European Southern Observatory, Karl-Schwarzschild-Str. 2, 85748 Garching bei München, Germany \\
$^{7}$INAF – Osservatorio Astrofisico di Arcetri, Largo E. Fermi 5, I-50125 Firenze, Italy\\
}

\date{Accepted XXX. Received YYY; in original form ZZZ}

\pubyear{2022}

\begin{document}
\label{firstpage}
\pagerange{\pageref{firstpage}--\pageref{lastpage}}
\maketitle

\begin{abstract}
The mechanisms that maintain turbulence in the interstellar medium
(ISM) are still not identified.
This work investigates how we can distinguish between two fundamental driving mechanisms: the accumulated effect of stellar feedback versus the energy injection from Galactic scales.
We perform a series of numerical simulations describing a stratified star forming ISM subject to self-consistent stellar feedback.
Large scale external turbulent driving of various intensities is added to mimic galactic driving mechanisms.
We analyse the resulting column density maps with a technique called Multi-scale non-Gaussian segmentation that separates the coherent structures and the Gaussian background.
This effectively discriminates between the various simulations and is a promising method to understand the ISM structure.
In particular the power spectrum of the coherent structures flattens above 60 pc when turbulence is driven only by stellar feedback.
When large-scale driving is applied, the turn-over shifts to larger scales.
A systematic comparison with the Large Magellanic Cloud (LMC)
is then performed.
Only 1 out of 25 regions has a coherent power spectrum which is consistent with the feedback-only simulation.
A detailed study of the turn-over scale leads us to conclude that regular stellar feedback is not enough to explain the observed ISM structure on scales larger than 60 pc.
Extreme feedback in the form of supergiant shells likely plays an important role but cannot explain all the regions of the LMC. If we assume ISM structure is generated by turbulence, another large scale driving mechanism is needed to explain the entirety of the observations.
\end{abstract}




\section{Introduction}

It has become clear that turbulence plays an important role in the star formation process \citep[see, e.g., the reviews of ][]{elmegreen04, maclow04, mckee07, hennebelle12, klessen16}.
Compressible supersonic turbulence generates a complex network of shocks and filaments as seen in observation of the interstellar medium (ISM).
Because turbulence naturally dissipates on small scales, a form of driving at larger scales is needed to maintain the turbulent kinetic energy in the system.
The regime between the injection scale and the dissipation scale is called the inertial range and contains fully developed turbulence \cite[see, e.g.,][]{lesieur08}.
Stellar feedback in the form of supernovae, jets, winds and ionizing radiation has been shown to be an important driving mechanism \citep{deavillez2005,Joung_MacLow2006,Kim_et_al2013, Girichidis_et_al2016, Padoan_et_al2016, Iffrig_et_al2017}.
These processes shape the structure of the ISM and can sometimes completely destroy the molecular clouds that are the formation sites of the stars.

Bubbles blown by the feedback of a single massive star have typical sizes of a few tens of parsecs.
When multiple massive stars form in close proximity, bubbles can combine to form structures with sizes of the order of 100~pc \citep{Chu2008}.
Stellar feedback injects energy and momentum into the ISM on the scales of individual stars and their feedback bubbles \citep{Walch_et_al2015, Girichidis_et_al2016}. 
Another mechanism for driving turbulence, which is much less investigated, is the energy cascade from larger scales, i.e. the galactic scale, to the molecular cloud scale \citep{klessen10}.
It has been shown that global galactic motions can indeed drive the turbulence on scales larger than 1 kpc \citep{wada2008,Bournaud_et_al2010, Renaud_et_al2012, Krumholz_et_al2018, Meidt_et_al2018, Nusser_et_Silk2021}.
The energy injected on these large scales then cascades further down and provides turbulent energy on smaller scales.
Although the properties of turbulence in the inertial range do not depend on the injection mechanism, we can still expect these different driving mechanisms to alter the structure and properties of the ISM in different ways.
In particular, the scales which are of the order of the injection scale can carry signatures of the driving mechanism.

It is currently unknown how important large-scale driving is for star formation and what imprint it leaves on the structure of the ISM.
In a previous study, \cite{Brucy_et_al2020} investigated the effect of external driving on the star formation rate (SFR) in the context of high column density regions.
These simulations showed that strong external driving is needed to quench the star formation and obtain the correct slope for the observed Kennicutt-Schmidt relation \citep{Kennicutt&Evans2012}, which correlates the SFR with the gas column density of the star forming region.
In this work, we study another aspect of large-scale driving, i.e.\ the structure of the ISM.
We focus on the global structure in systems with characteristics typical of present-day galaxies.
The properties of substructures such as molecular clouds and star clusters will be addressed in a future paper.

For the purpose of this study, we run a suite of simulations representing a 1 kpc part of a galactic disk.
The simulations, which are presented in detail in Section~\ref{sec:setup}, include a large range of physics including stellar feedback and various external driving strengths.
The external driving is applied on scales of a 1/3, 1/2 and a full box length, significantly larger than the size of a typical stellar feedback bubble.
We analyse the final ISM structure and look for observable signatures of large-scale driving.
The tool we use in this work is called Multi-scale non-Gaussian segmentation (MnGseg) and is described in \citet{Robitaille_et_al2014,Robitaille_et_al2019} as well as in Section~\ref{sec_MnGseg_theory}.
According to \cite{Falgarone_et_al2004}, the ISM can be decomposed into two components: a Gaussian fluffy fractal component and a component containing the dense structures such as cores and filaments.
MnGseg can separate these two components, dubbed Gaussian and coherent, allowing us to study their individual characteristics. 
The statistic we are particularly interested in is the power spectrum (PS) which, simply put, gives us a measure of how much structure there is as a function of scale.
Indeed, the column density PS has been used in many observational and theoretical papers to characterize the ISM \citep[see, e.g.,][]{Miville-Deschenes_et_al2007,Kritsuk_et_al2007, federrath10, federrath13, schneider15}.
In this work, rather than measuring the classical Fourier, MnGseg uses a wavelet-based method.
It turns out that the PS of the separated components (Gaussian and coherent) carries, for our purpose, more information than the total PS, as is demonstrated in Section~\ref{sec:MnGseg_sims} where we apply this technique to the simulation results.
Because MnGseg has never been applied to simulations before, we perform a series of tests which helps us to determine what algorithm parameters are most suited, how to interpret the results and how we can best compare with observation.

Once we have identified the signature of large-scale driving on the power spectrum decomposition, we can look for it in observations.
For this, we apply MnGseg to Herschel 500 $\mu$m observations of the Large Magellanic Cloud (LMC).
We need an external galaxy for this type of analysis and the LMC provides the best target since it is close-by and face-on, which means we can study a large range of scales.
The results of this are described in Section~\ref{sec:MnGseg_LMC}.
The statistical properties of the LMC have been studied by many authors \citep[e.g.,][]{Elmegreen_et_al2001, Szotkowski_et_al2019, Koch_et_al2020}.
However, a decomposition as done by MnGseg has not yet been attempted.
In fact, so far MnGseg has only been applied on regions much smaller than 1 kpc.
Using the insights gained from the simulations as means to interpret the observations, we determine whether large-scale driving might play an important role.

Finally, in section \ref{sec:discussion}, we discuss the results and caveats.
After obtaining clues from local variation within the LMC, we speculate about the possible sources of large-scale driving.
Note that the LMC is not only convenient because it is close-by.
It also has interesting features which are candidates for large-scale driving sources: it has large-scale stellar structures in the form of spiral arm(s) and a bar, and it is tidally interacting with the Milky Way and Small Magellanic Cloud (SMC).
If we find evidence that any of these can inject turbulence from large scales into the ISM of the LMC, we could expect them to play a role in other galaxies as well.
This, of course, should then be verified by dedicated studies of other galaxies.

We end the paper with a summary of our conclusions.

\section{Simulation setup}
\label{sec:setup}

The simulation setup is similar to the ones used in previous works \citep{Iffrig&Hennebelle2015, Colling_et_al2018, Brucy_et_al2020}, however in this study we allow for deeper adaptive mesh refinement up to a minimum cell size of 0.24 pc.
This allows us to more accurately study the morphological and dynamical properties of dense interstellar structures.

\subsection{Initial conditions}

We model a 1 kpc part of a galactic disk.
Initially, the density profile along the $z$-axis is given by a Gaussian
\begin{equation}
n(z) = n_0 \exp{ \left[ - \frac{1}{2} \left( \frac{z}{z_0} \right)^2 \right]}
\end{equation}
with $n_0$ the mid-plane density and $z_0$ the thickness of the disk.
We adopt $n_0 = 1.5 \, \mathrm{cm}^{-3}$ and $z_0 = 150$ pc corresponding to a gas column density of 19.1 \Msun $\mathrm{pc}^{-2}$, a value which slightly higher than the average value found in the LMC but in good agreement with dense sub-regions (as can be inferred from the data presented in \citet{kim_et_al2003, Wong_et_al2011}).
This profile is embedded in an external gravitational potential of the form
\begin{equation}
g(z) = - \frac{a_1 z}{\sqrt{z^2+z_0^2}} - a_2 z
\end{equation}
with $a_1 = 1.42 \times 10^{-3}$ kpc Myr$^{-2}$,
$a_2 = 5.49 \times 10^{-4} \, \mathrm{Myr}^{-2}$ and
$z_0 = 0.18$ kpc
\citep{Kuijken_Gilmore1989, Joung_MacLow2006}.
Self-gravity is also considered.
An initial level of turbulence is introduced by adding a turbulent velocity field with a root mean square dispersion of 5 km s$^{-1}$ and a Kolmogorov power spectrum $E(k) \propto k^{-5/3}$ with random phase.
The initial temperature is 5333 K, which is a typical value for the warm neutral medium (WNM) phase in the ISM.
We also include an initial Gaussian magnetic field with an orientation along the $x$-axis
\begin{equation}
B_x(z) = B_0 \exp{ \left[- \frac{1}{2} \left( \frac{z}{z_0} \right)^2 \right]}
\end{equation}
with $B_0 = 7.62 \, \mu$G, comparable to the field strength in the Milky Way and LMC \citep{Gaensler_et_al2005, Hassani_et_al2021}.
The box is periodic in the $x$- and $y$-direction and has open boundary conditions in the $z$-direction.

\subsection{Numerics}

To evolve our simulation in time, we use the ISM version of the adaptive mesh refinement hydrodynamics code \textsc{ramses} \citep{Teyssier2002}, with treatment of the magnetic field using ideal MHD \citep{Fromang_et_al2006} and radiation using the M1 method \citep{Rosdahl_et_al2013}.
Cooling is as described in \cite{Audit_Hennebelle2005}.
We also include heating from a uniform UV background with strength equal to the solar neighbourhood field.

The coarse grid has a resolution of 3.9 pc (refinement level\footnote{When refined, the cell is divided by 2 along each spatial axis. The refinement level indicates how many times this division is done.} 8).
The grid is then refined further when a cell exceeds a certain mass\footnote{The thresholds are $5.7 \times 10^{-4}, 7.1 \times 10^{-4}, 4.4\times 10^{-4},$ and $1.1 \times 10^{-4}$ \Msun.}, up to level 12 corresponding to a maximum resolution of 0.24 pc in the densest regions.
We introduce sink particles \citep{Bate_et_al1995, Federrath_et_al2010} when the gravitational collapse reaches the resolution limits, which mimics the formation of a star cluster and helps to prevent singularities in the computational domain.
We use the sink formation algorithm from \citep{Bleuler&Teyssier2014}.
Sinks are created from overdensities identified by the native \textsc{ramses} clumpfinder \citep{Bleuler_et_al2015} if their density exeeds the threshold of $10^4$ H cm$^{-3}$.
Additionally, this implementation checks that the gas clump from which the sink is forming is collapsing and bound.
After their birth, sinks accrete gas according to the threshold accretion scheme:
only gas which is above the sink formation threshold and within the accretion radius of 4 cells will be accreted, with a maximum of 75\% of the mass available per time step.
No additional checks are applied before the gas is accreted.
New sinks cannot form within the accretion radius of existing sinks.
Sinks are not allowed to merge.
All simulations have been run for 60 Myr during which 2-5 $\%$ (depending of the simulations) 
of the gas  has been converted into stars.
After this time, the turbulence is fully developed as shown in Appendix~\ref{sec:appendix_steady_state}.

\subsection{Stellar feedback}
Sink particles also serve as a source of stellar feedback.
Each time a sink has accreted a mass of 120 \Msun, a massive star particle with a mass randomly determined from the Salpeter IMF \citep{Salpeter1955} between 8 and 120 \Msun\ is created.
We associate a lifetime $\tau_*$ with this star using the model
\begin{equation}
    \tau_*(M) = \tau_0 \exp \Big[ - a  \Big(\log \Big(\frac{M}{M_0} \Big) \Big)^{b} \Big]
\end{equation}
\noindent with $\tau_0=3.265$ Myr, $M_0=148.16$ \Msun, $a=0.238$, $b=2.205$ \citep{Woosley_et_al2002}.
Once this massive star has reached the end of this lifetime,
it explodes in a random location within a sphere of radius $\tau_* \times$ 1 km s$^{-1}$.
Given the typical lifetimes of massive stars, this can range from a fraction of a parsec to about 30 pc.
The explosion injects a momentum of roughly $4 \times 10^{43}$ g cm s$^{-1}$ (but no thermal energy) into the ISM \citep{Iffrig&Hennebelle2015}, mimicking a supernova.

We also include self-consistent feedback from HII regions, with energy and momentum injected according to the flux of ionising photons emitted by the star.
The evolution of HII regions itself is computed by the radiative transfer module of \textsc{ramses} \citep{Rosdahl_et_al2013}, as in \cite{Geen_et_al2016}.
For details about the exact implementation, see \cite{Colling_et_al2018} and references therein.

Due to their computational cost, we do not include stellar winds.

\subsection{External driving}

In this work, we study the effect of turbulent energy injection from large scales.
A turbulence driving force is added as an additional external force in the Euler equation.
The Fourier modes of this force are computed using the generalised Ornstein-Uhlenbeck process and follow a stochastic differential equation:
\begin{equation}
\label{eq:dF}
    \mathrm{d} \vec{F} (\vec{k},t) = F_0 (\vec{k}) \vec{P}\left(\begin{array}{c} k_x \\ k_y \\ 0\end{array}\right) \mathrm{d} \vec{W}_t - \vec{F}(\vec{k},t) \dfrac{\mathrm{d} t}{T}
\end{equation}
\citep{Eswaran_Pope1988, Schmidt_et_al2006, Schmidt_et_al2009}.
The first term describes a stochastic contribution to the force.
$\vec{F}$ is a complex 3D vector in Fourier space.
$F_0$ is the power spectrum assigning a weight to each $\vec{k}$-mode.
In this study, we drive on scales between 1 box length and 1/3 of the box length with a parabolic power spectrum
\begin{equation}
    \label{eq:F0}
    F_0(\vec{k}) = 
    \begin{cases} 
    1 - \left(\vert k \vert - 2\right)^2\text{ if } 1 < \vert k \vert < 3 \text{ and } k_x > 0 \text{ and } k_y > 0\\
    0 \text{ otherwise}
    \end{cases}
\end{equation}
which peaks at scales of half the box size. The purely vertical modes are removed.
$\vec{P}(\vec{k})$ is the projection operator which takes care of compressive versus solenoidal modes through a Helmholtz decomposition.
It is defined as
\begin{equation}
    \vec{P}(\vec{k}) =  \zeta \vec{P}^{\perp}(\vec{k}) + (1 - \zeta) \vec{P}^{\parallel}(\vec{k}) 
\end{equation}
with $\vec{P}^{\perp}$ and $\vec{P}^{\parallel}$ the projection operators respectively perpendicular and parallel to $\vec{k}$ \citep{Federrath_et_al2010}.
For the solenoidal fraction $\zeta$, we choose $\zeta=0.75$.
This is higher than the standard value of 0.5 for a natural mix \citep{Federrath_et_el2008}.
However, \cite{Jin_et_al2017} find that in their galaxy simulations the velocity field is mainly solenoidal which may hint that turbulence driving by galactic dynamics might be dominantly solenoidal.
Because our setup is a stratified disk, the projection is done onto the 2D plane of the disk (the projection operator $\vec{P}$ is applied to $(k_x, k_y, 0)$ instead of $\vec{k}$ in Equation \ref{eq:dF}).
d$\vec{W_t}$ is the Wiener process, which gives us a random vector from a Gaussian distribution with zero mean and variance.
d$t$ is the timestep of integration.
The second term is an exponential decay with $T$ the autocorrelation time, which is set to $T=40$ Myr.

Once the Fourier modes $\vec{F}$ are calculated, they are converted into a real force by applying a Fourier transform. 
The force is then multiplied by a boost factor $f_{\mathrm{rms}}$ to adjust its strength, measured through the time average of the root mean square (RMS) value of the Fourier coefficients:
\begin{equation}
    \label{eq:rms}
    \mathrm{RMS} = \left< f_{\mathrm{rms}}  \sqrt{\int\left\vert\vec{F}(\vec{k},t)\right\vert^2  \mathrm{d}^3\vec{k} }\right>_t,
\end{equation}
where  $\left<\cdot\right>_t$ denotes the time average.

The parameters that characterize the turbulence driving are the boost factor (which sets the RMS value of Equation \ref{eq:rms}), the auto-correlation time, the power spectrum of the modes and the solenoidal fraction.
In this study, we vary the driving strength while keeping the other parameters constant.
We take boost factors $f_{\mathrm{rms}}$ of 3000, 6000 and 24000, corresponding to weak, medium and strong driving respectively.
For completeness, the corresponding values for the RMS are listed in Table~\ref{tab:energies}.
These parameters values correspond to three different final velocity dispersions and have been chosen through experimentation.
The energy injection for each driving strength is listed in Table~\ref{tab:energies} and will be discussed in Section~\ref{sec:energy_estimates}.
We also run a simulation without driving.
This allows us to separate the effects of driving by stellar feedback from those of external large-scale driving.

\section{Multi-scale non-Gaussian segmentation}
\label{sec_MnGseg_theory}

The MnGseg analysis was developed by \cite{Robitaille_et_al2014, Robitaille_et_al2019} and is inspired by the analysis of turbulent fluid flows of \cite{2012PhyD..241..186N}.
The implementation is freely available as the python package \textsc{pywavan}\footnote{\url{https://github.com/jfrob27/pywavan}}.
For a detailed prescription we refer to the above mentioned works.
Here, a short summary is given. For the sake of readability and conciseness most of the technical details and several illustrative figures are provided in Appendix~\ref{sec:mngseg_appendix}.

\subsection{Wavelet power spectrum}
Structures with a certain size and orientation can be identified in an image $f(\vec{x})$ by  convolving it with a complex Morlet wavelet of size $\ell$ and orientation angle $\theta$ \citep{Robitaille_et_al2014}.
The convolution operation for the wavelet transform is done in the Fourier space, 

\begin{equation}
\label{eq:wavelet_coeff}
\tilde{f}_{\ell, \theta}(\vec{x}) = \mathcal{F}^{-1}\left\{\hat{f}(\vec{k})\hat{\psi}_{l,\theta}^*(\vec{k})\right\}
\end{equation}

\noindent where $\mathcal{F}^{-1}$ denotes the inverse Fourier transform, and $\hat{f}(\vec{k})$ and $\hat{\psi}_{\ell,\theta}(\vec{k})$ denote respectively the Fourier transform of the image and of the Morlet wavelet $\psi_{\ell,\theta}(\vec{x})$.
Figure~\ref{fig:how_mnGseg_Morlet} illustrates the resulting wavelet coefficients $\tilde{f}_{\ell, \theta}$ for three different sizes and angles.
To calculate the total wavelet power spectrum $P(\ell)$, one simply averages the squared absolute value of the complex wavelet coefficients over angles (Figure~\ref{fig:how_mnGseg_average_angles}) and then over spatial coordinates $\vec{x}$:

\begin{equation}
P(\ell) = \langle|\tilde{f}_{\ell,\theta}(\vec{x})|^2\rangle_{\theta,\vec{x}}
\label{eq:wavelet_powspec}
\end{equation}

\noindent as illustrated in Figure~\ref{fig:how_mnGseg_image_reconstruction}.
Equation \ref{eq:wavelet_powspec} calculates the second-order moments of wavelet coefficients. It can be shown that using the complex Morlet wavelet, this measurement becomes equivalent to the standard Fourier power spectrum analysis \citep{Robitaille_et_al2019, 2005CG.....31..846K} as can be seen in Figure~\ref{fig:how_mnGseg_image_reconstruction}.

We can estimate the error on the wavelet PS in a way similar to what is done for the Fourier PS.
The number of times a wavelet of size $\ell$ fits into the domain sets the number of samples $S(\ell) = L / \ell$ for that scales, where $L$ is the size of the full domain.
The larger $\ell$, the smaller $S$ and the larger the error.
Analogous to the Fourier PS, we define the error due to limited sampling as
\begin{equation}
    \mathrm{error}(\ell) = \frac{P(\ell)}{S(\ell)}
\end{equation}
\noindent For a domain which spans 1 kpc, this results in an error of less than 20\% for scales below 200 pc.

\subsection{Decomposition into Gaussian and coherent part}

Once we have the convolved images, through an iterative process as a function of $\ell$ and $\theta$, MnGseg separates the Gaussian and coherent component.
The fractal Gaussian part is, as its name suggests, associated with a Gaussian signal in the image.
Plotting the histogram of wavelet coefficients $\tilde{f}_{\ell}(\vec{x})$ of an originally fractal image results in self-similar Gaussian distributions at every scale. Note that depending on the power law of $P(\ell)$, the total image with all integrated spatial scales is not necessary Gaussian.
In complex systems like the ISM, many physical processes shape the structure, introducing clumps, filaments and bubbles. The presence of these `coherent' structures will alter the distribution by adding a non-Gaussian component.
In other words, if one can identify and remove the Gaussian component, whatever remains are coherent structures.
The procedure MnGseg follows to separate the two components is to count the high intensity tail in the distribution as coherent.
This segmentation is illustrated in Figure~\ref{fig:how_mnGseg_segmentation_histogram}.

The segmentation procedure introduces a segmentation parameter $q$ to regulate how strict we want our Gaussian distribution to be.
Gaussianity is evaluated on the absolute value of the complex Morlet wavelet coefficients $|\tilde{f}_{\ell,\theta}(\vec{x})|$.
Consequently the coefficient distributions correspond to Rician distributions, rather than a pure Gaussians, and a small skewness is expected.
This $q$ parameter is involved in the iterative process used to converge to the best threshold $\Phi$ \citep{2012PhyD..241..186N}:

\begin{equation}
\left\{
        	 \begin{array}{l}
  	   \Phi_{0}(l,\theta)=\infty \\
 	   \Phi_{n+1}(l,\theta)=q \, \sigma_{l,\theta}(\Phi_{n}), \\
 	 \end{array}
  	 \right.
\label{eq:q_parameter}
\end{equation}

\noindent where $\sigma_{l,\theta}(\Phi_{n})$ is defined as,

\begin{equation}
\sigma_{l,\theta}^2(\Phi)=\frac{1}{N_{l,\theta}(\Phi)}\sum_{\vec{x}} \mathbb{L}_{\Phi} \big( |\tilde{f}_{l,\theta}(\vec{x})| \big) \, |\tilde{f}_{l,\theta}(\vec{x})|^2,
\end{equation}

\noindent{with}

\begin{equation}
\mathbb{L}_{\Phi} \big(|\tilde{f}_{l,\theta}(\vec{x})|\big)=\left\{
       									          \begin{array}{ll}
  									          1 & \qquad \mathrm{if} \quad |\tilde{f}_{l,\theta}(\vec{x})| < \Phi \\
 									          0 & \qquad \mathrm{else}, \\
 									         \end{array}
  									         \right.
\end{equation}

\noindent{and}

\begin{equation}
N_{l,\theta}(\Phi)=\sum_{\vec{x}} \mathbb{L}_{\Phi} \big( |\tilde{f}_{l,\theta}(\vec{x})| \big).
\end{equation}
\noindent
Choosing a large value for $q$ will result in a large fraction of the signal being counted as Gaussian, as the cut-off for coherent structures shifts to larger, more extreme values of $|\tilde{f}_{l,\theta}|$.
On the other hand, choosing a small $q$ will categorize almost all values of $|\tilde{f}_{l,\theta}|$ as coherent structure.
MnGseg can optimise the dimensionless parameter $q$ for each individual scale by measuring the skewness of the coefficient distributions and iterating until a desired skewness is obtained.
After many experiments, we settled on a fixed $q=2$, a value in line with what is expected from theory \citep{2012PhyD..241..186N} and comparable to what is used in previous studies (\citealt{Robitaille_et_al2019}, Cunningham et al. in prep.).
Where instructive, we show results obtained with a different values.
The segmentation obviously depends on the adopted value of $q$.
However, when we apply MnGseg on several different data sets using the same $q$ for all data, the resulting relative differences between data sets remains largely independent on the adopted value of $q$.
This seemingly indicates the validity and robustness
of the conclusions (see Appendix~\ref{sec:test_q} and the right column of Figure~\ref{fig:powerspectra_all}).

\subsection{Application to simulations}
Whereas in principle the segmentation can be performed on the full 3D data cube, the observations we want to compare to are inherently 2D projections onto the sky, so we will limit ourselves to analysing projections of the density field of our simulations.
The face-on projection lends itself excellently to this task, since it has periodic boundaries, avoiding any artifacts that may result from Fourier transforming a non-periodic image. 
To analyse the edge-on views, we could select only a part of the disk.
These regions would be of a size determined by the thickness of the disk, which is about 250 pc.
However, since the external driving we chose in our setup operates on scales between 1 kpc and 333 pc, its signature is likely to be difficult to capture in images of regions smaller than the injection scale.
Another reason to focus on the face-on views is that it is fairly straightforward to compare with face-on external galaxies.
When looking at edge-on galaxies, the sight line is much larger than the 1 kpc which is modelled here.

To generate the column density images, we use a wrapper around the \textsc{amr2map} tool provided with \textsc{ramses}.
The density is projected along the $z$-axis of the simulation grid onto an array of 1024 $\times$ 1024 pixels.
This means that one level 10 cell corresponds to 1 pixel.
Contributions from cells which are at higher refinement levels are weighted by their size.
 
\section{Simulation results}
\label{sec:MnGseg_sims}

\subsection{Determining the signature of large-scale driving}

\begin{figure*}
    \centering
    \includegraphics[width=\textwidth]{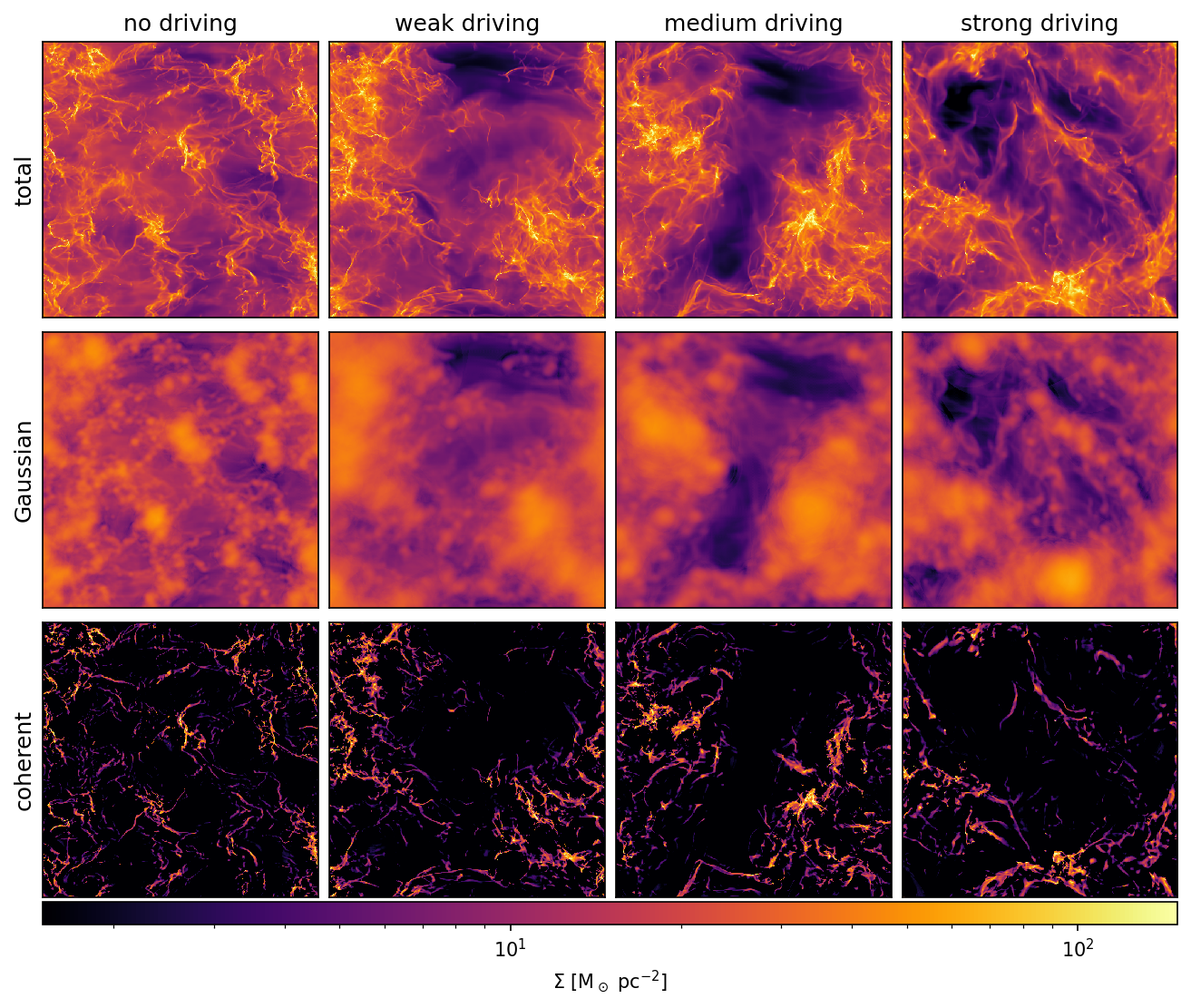}
    \caption{MnGseg applied on the face-on view at the end of the simulations ($t_{\mathrm{end}}=$ 60 Myr). Each images spans the full 1 kpc$^2$ of the simulation box. Top: the original image, middle: Gaussian component of the image (including average), bottom: coherent component (excluding average) of the image. The total image is the sum of the Gaussian and the coherent part.}
    \label{fig:images_decomposition}
\end{figure*}

\begin{figure*}
    \centering
    \includegraphics[width=\textwidth]{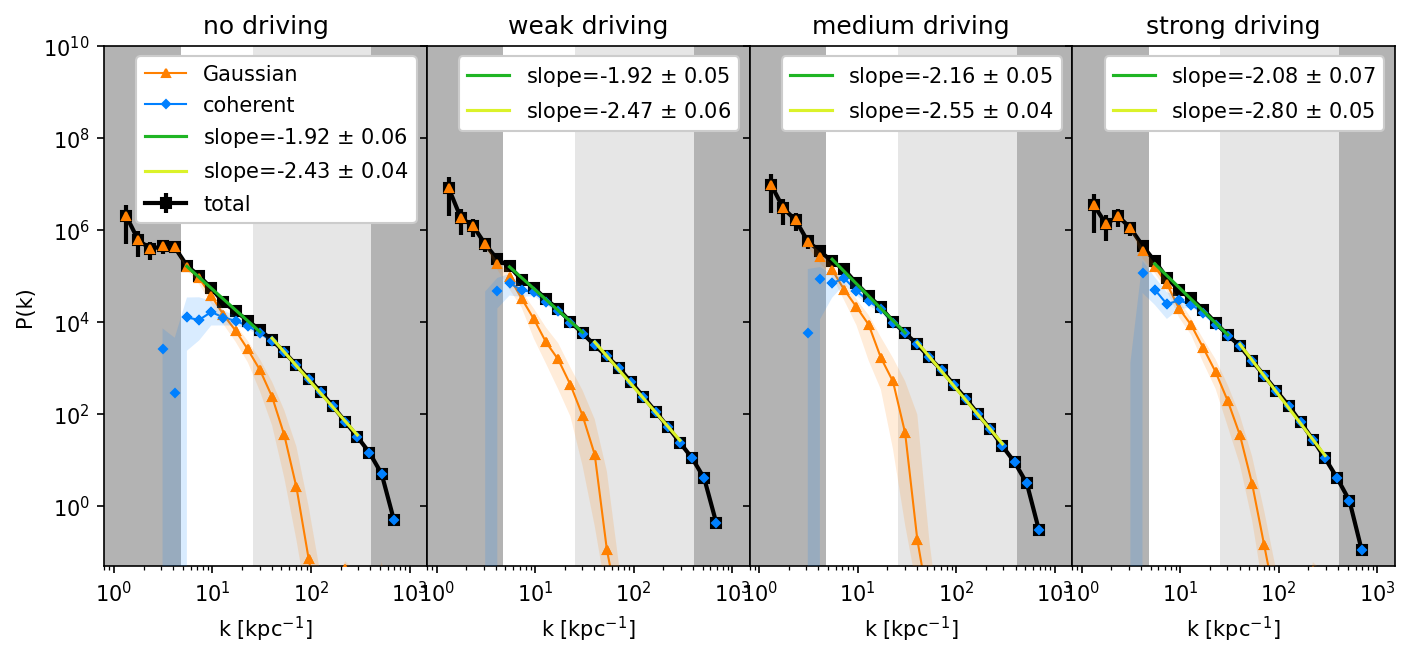}
    \caption{The decomposition of the power spectra into Gaussian and coherent part, corresponding to the density maps of Fig.~\ref{fig:images_decomposition} ($q=2.0$). The green lines show power-law fits to the total PS. The grey areas mark regions where uncertainties are large due to limited statistics (small $k$) and limited resolution of the simulation (large $k$) where the light and dark grey indicates 10 times the coarse grid and fine grid resolution, respectively.
    The colored bands show the segmentation results when varying $q$ between 1.8 and 2.2.}
    \label{fig:powerspectra_decomposition}
\end{figure*}

\begin{figure*}
    \centering
    \includegraphics[width=\textwidth]{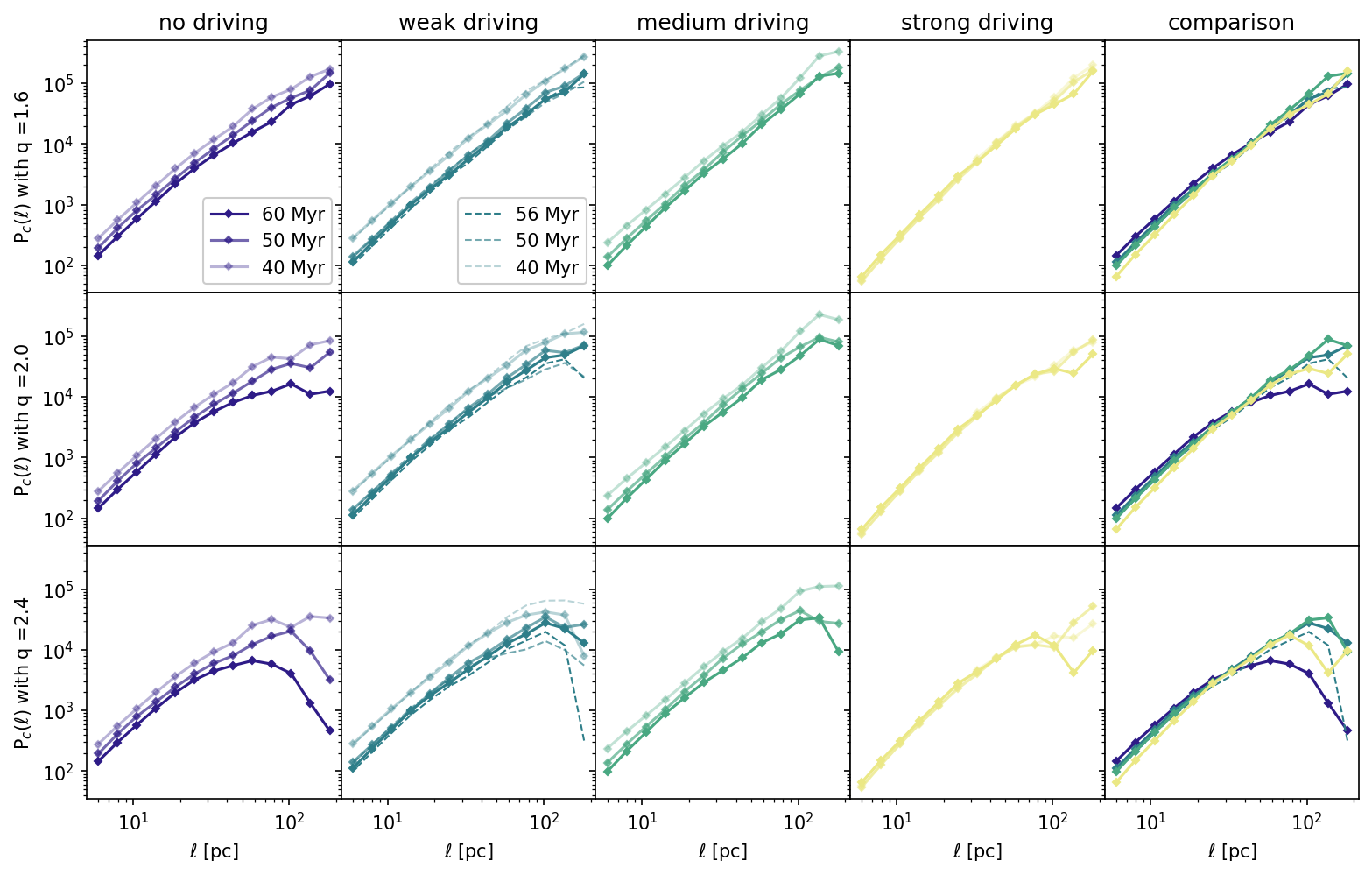}
    \caption{Variation of the coherent part of the PS with segmentation parameter $q$ and time. The dashed line shows the result of a weak driving simulation with a different turbulence driving seed. The right-most column compares the results at the end of the simulations (60 Myr).}
    \label{fig:powerspectra_all}
\end{figure*}

We now apply MnGseg to the face-on view of our simulations.
Figure~\ref{fig:images_decomposition} displays the resulting image decomposition.
We see the overdense filaments and clumps are well recovered in the coherent image, whereas the Gaussian image looks like a fluffy fractal cloud, as expected.
The corresponding power spectra are shown in Figure~\ref{fig:powerspectra_decomposition}.
The total PS can be described by two power laws: one for large scales in which the diffuse part is well resolved and one for smaller scales which probe the adaptive mesh refinement region.
We fit the slope in both of these regimes.
The total PS for the no-driving case and the weak driving case is the same, while in the medium and strong driving case it is slightly steeper.
For small scales, there is a trend of steeper slope with increasing driving strength.
A similar trend may be present in for large scales, though it is less clear.

When we do the decomposition, the Gaussian component dominates the large scales (small $k$) and sharply drops at small scales, due to the lack of resolution in the diffuse gas in the simulation, since the mesh is refined only in the dense regions.
Comparing the Gaussian PS to the resolution limit for diffuse gas, i.e. 10 times the coarse grid cell size indicated in light grey in the figure, seems to confirm this hypothesis.
The coherent part becomes the dominant contribution above a turn-over $k_{\mathrm{LS}}$.
The associated scale could be interpreted as a maximum coherent structure size which the turbulent driving is able to generate.
A clear distinction can be made between the simulations with and without external driving.
For the no-driving case, the coherent PS flattens at large scales.
In the simulations with driving, the power law continues towards larger scales with a turn-over close to the statistics limit. 
We define the turn-over scale $\ell_{\mathrm{LS}}$ quantitatively as the largest scale where $P_\mathrm{coherent} > P_\mathrm{Gaussian}$.
Without driving, $\ell_{\mathrm{LS}} =$  58 pc.
In the weak and medium driving case, this turn-over is shifted to larger scales, with $\ell_{\mathrm{LS}}=$ 137 pc.
For the strong driving, we do not see a clear flattening of the coherent power spectrum, through there is a small dip at 103 pc, roughly at the same scales where the weak and medium coherent PS have their turn-over.

To verify these trends, we repeat the analysis for different times in the simulations and for different values of the segmentation parameter $q$.
A full study of the variations with $q$ can be found in Appendix~\ref{sec:test_q}.

The right column of Figure~\ref{fig:powerspectra_all} compares the coherent PS at the end of each simulation (60 Myr) for different values of $q$.
For $q=1.6$ it is difficult to distinguish the simulations,
since most of the signal is labeled as coherent.
When increasing $q$, we see the large scale coherent PS becomes weaker rapidly in the no-driving case.
The results for simulations with driving are much less affected by the change in $q$
This means it is more difficult to recover large scale coherent structure in the no-driving case compared to the simulations with driving.
The dashed line shows the result for a simulation with weak driving and a different turbulence driving seed.
This reassures us that the location of the turn-over is not a statistical fluctuation caused by the random driving.

One thing that requires some explanation is the dip for the strong driving case, which becomes even more pronounced when increasing $q$ as seen in Figure~\ref{fig:powerspectra_all}.
To gain some insight, we look at the time evolution.
The left part of this figure shows the coherent PS at different times in each simulation.
What stands out is that in the no-driving case, there is a clear decrease at large scales over time, which is not observed in the simulations with external driving.
What we see here is the imprint of the initial conditions dissipating.
As explained in Section~\ref{sec:setup}, we start the simulation with a random turbulent velocity field.
This field has large scale modes, which dissipate as the simulation evolves.
In the simulations with external driving, these modes are replenished.
Without external driving, they decay.
When looking at the time evolution for the strong driving case, there is a a small increase in large scale power from 40 to 50 Myr, while the dip is developed between 50 and 60 Myr.
It is thus likely a statistical fluctuation of the large scale geometry.
This illustrates that the uncertainties on large scales are significant, but everything smaller than 100 pc is well-described.
This ensures the turn-over observed around 50 pc in the no-driving case is real and no such flattening is observed in the cases with driving until scales which are \textit{at least} twice as large.

In summary, a clear difference can be seen when comparing the coherent PS of the simulation without external driving to the ones with driving.
Whether there is a difference between the individual simulations with driving is debatable.
There might be an indication that stronger driving leads to steeper power spectra, but more data is needed to draw robust conclusions.
Overall this PS analysis confirms what can be seen visually in the column density maps: large scale external turbulence driving generates large scale coherent structure in the ISM. The typical size scale of these structures is larger than what can be generated with stellar feedback alone.

\subsection{Estimates of the injected energy}
\label{sec:energy_estimates} 

\begin{table*}
    \centering
    \begin{tabular}{r c c c c c c c}
        driving & $f_{\mathrm{rms}}$ & RMS ({\footnotesize km s$^{-1}$ Myr$^{-1}$})& $E_{\mathrm{turb}}$ [$10^{51}$ erg] & $E_{\mathrm{SN}}$ [$10^{51}$ erg] ({\footnotesize $\varepsilon=$ 2 - 10 \%}) & $\sigma_\mathrm{3D}(t=t_{\mathrm{end}})$ [km s$^{-1}$]\\
        \hline
        none & 0 &  0 & -- & 40 - 204 & 8.49\\
        weak & 3000 & 0.27 & 144 & 50 - 254& 9.08\\
        medium & 6000 & 0.54 & 578 & 66 - 329 & 12.06\\
        strong & 24000 & 2.08 & 5905 & 46 - 229 & 20.13
    \end{tabular}
    \caption{Estimates of the total energy injected by SN and external turbulence driving during each simulation. $E_{\mathrm{SN}}$ is estimated as the number of supernova times $10^{51}$ erg times an efficiency $\varepsilon$. The definitions of $f_{\mathrm{rms}}$ and RMS are given in Eq.~\ref{eq:rms}. The last column shows the final mass-weighted velocity dispersion within the disk (defined as the region within 200 pc above and below the midplane).}
    \label{tab:energies}
\end{table*}

One obvious question at this stage is:
how much energy is injected from stellar feedback and
from external driving?

To answer this question, we compare the energy injected by external turbulent driving to the energy injected by supernova over the course of the simulation.
These numbers are shown in Table~\ref{tab:energies}.
The energy injected by external driving is directly  computed in the simulations, while the energy from the supernovae is simply obtained by multiplying their number by 10$^{51}$ erg, the energy injected by one supernovae, and an efficiency $\varepsilon$.
The efficiency of supernova
driving turns out to be rather low. \citet{Iffrig_et_al2017}, using an analytical model of the galactic scale-height, have estimated this to be on the order of few percents, possibly as low as 1-2$\%$.
This is likely due to several not exclusive facts: \textit{i)} supernovae efficiently dissipate their energy;
\textit{ii)} when they explode in a group or at high latitude, a significant fraction of their 
energy goes into galactic winds;
\textit{iii)} supernova explosions have proven to be rather inefficient in delivering kinetic energy to dense gas \citep{Iffrig&Hennebelle2015}.
Therefore multiplying $E_\mathrm{SN}$ by the corresponding efficiency, we see that in the weak driving case the injected turbulent energy, $E_\mathrm{turb}$ is likely slightly lower or comparible to $E_\mathrm{SN}$ while in the medium driving one, it is slightly higher.
This is in good agreement with the modest difference ($\simeq 25 \%$) found for the SFR (which will be presented in a future study) between the no driving and weak driving runs and somewhat larger one ($100 \%$) found between the no-driving and medium driving runs. 
On the other hand the energy injected in the strong driving case is roughly 10 times larger than in the medium driving case.
While this seems a very substantial difference, let us recall that the turbulent energy dissipation is proportional to $v_\mathrm{rms}^3$, implying that the expected rms velocity difference may only be of the order of $\simeq$2, as confirmed by the measure of the mass-weighted velocity dispersion (Table~\ref{tab:energies}).
This may be the reason why the coherent PS of the various driven 
runs present only limited differences. 

\section{Comparison to observations: the case of the LMC}
\label{sec:MnGseg_LMC}

\begin{figure}
    \centering
    \includegraphics[width=\columnwidth]{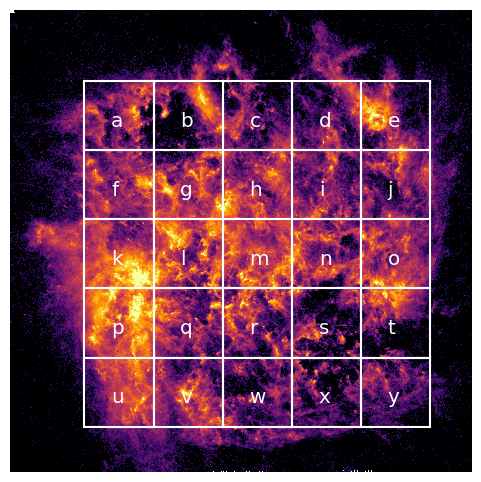}
    \caption{Image of the LMC as seen through Herschel at 500 $\mu$m and post-processed by \citet{Gordon_et_al2014}. We mark a series of 1 kpc$^2$ regions used to compare to the simulations.}
    \label{fig:LMC_cuts}
\end{figure}

\begin{figure}
    \centering
    \includegraphics[width=\columnwidth]{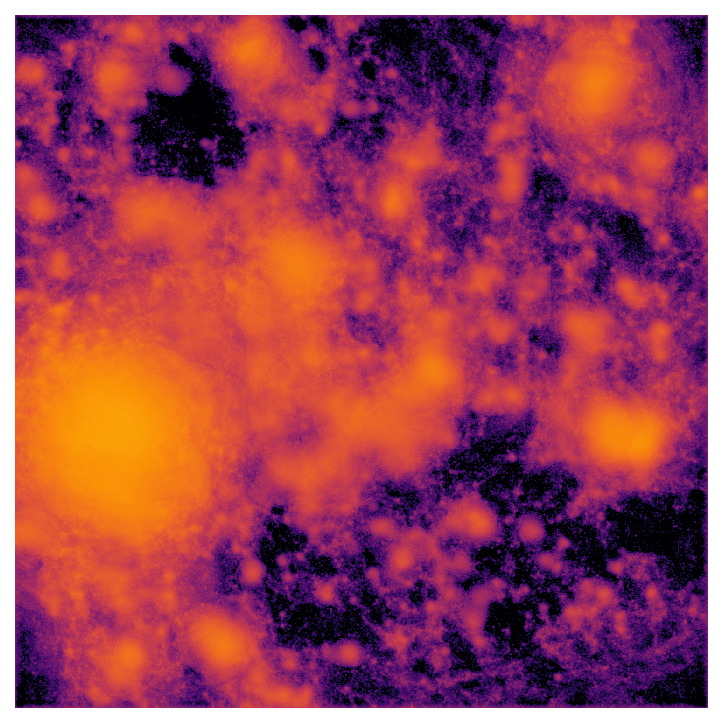}\\
    \includegraphics[width=\columnwidth]{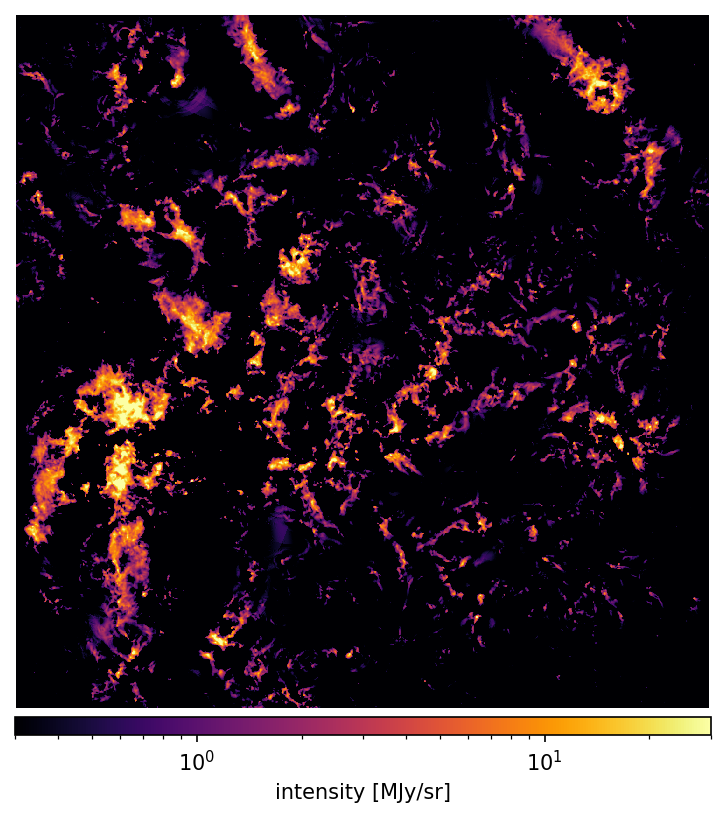}
    \caption{Image decomposition applied on the full 5 kpc $\times$ 5 kpc area of the LMC using $q=2.0$. Top: Gaussian component, bottom: coherent component.}
    \label{fig:image_decomp_LMC}
\end{figure}

\begin{figure*}
    \centering
    \includegraphics[width=\textwidth]{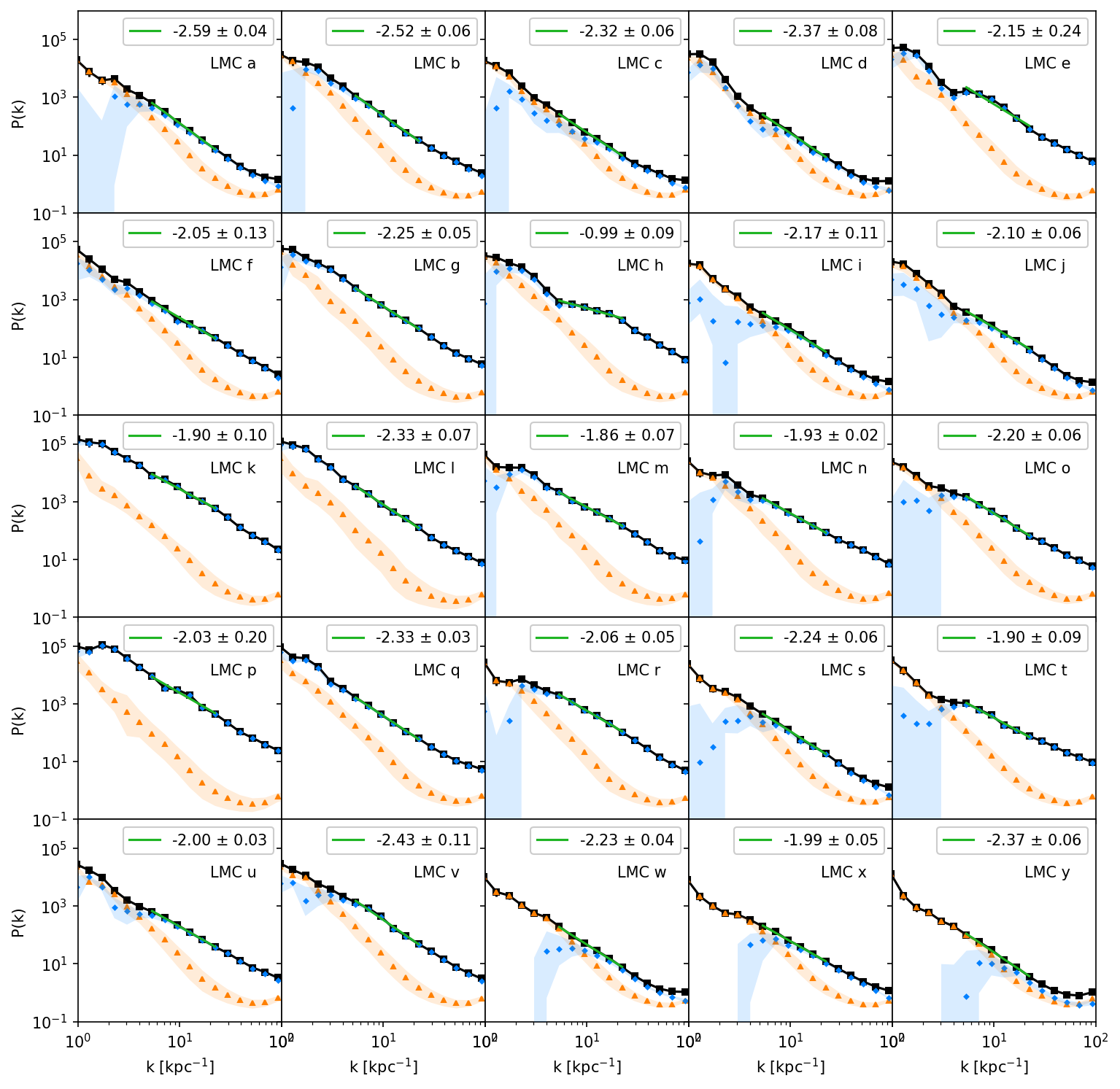}
    \caption{Power spectrum and its decomposition into Gaussian and coherent part with $q=2.0$ for a series of 25 different 1 kpc$^2$ region of the LMC, extracted from the full 5 kpc $\times$ 5 kpc analysis}. The green line shows a power law fit to the total PS in the same range as the upper slope measured in the simulations. On average total PS at 1 kpc has a slope of -2.13 $\pm$ 0.30. The error bars on the total PS are smaller than the points. The maximum value of $k$ is set by the beam size of the observation. The colored bands show the range for the segmentation when varying q between 1.8 and 2.2.
    \label{fig:LMC_PS_parts}
\end{figure*}

\begin{figure*}
    \centering
    \includegraphics[width=\textwidth]{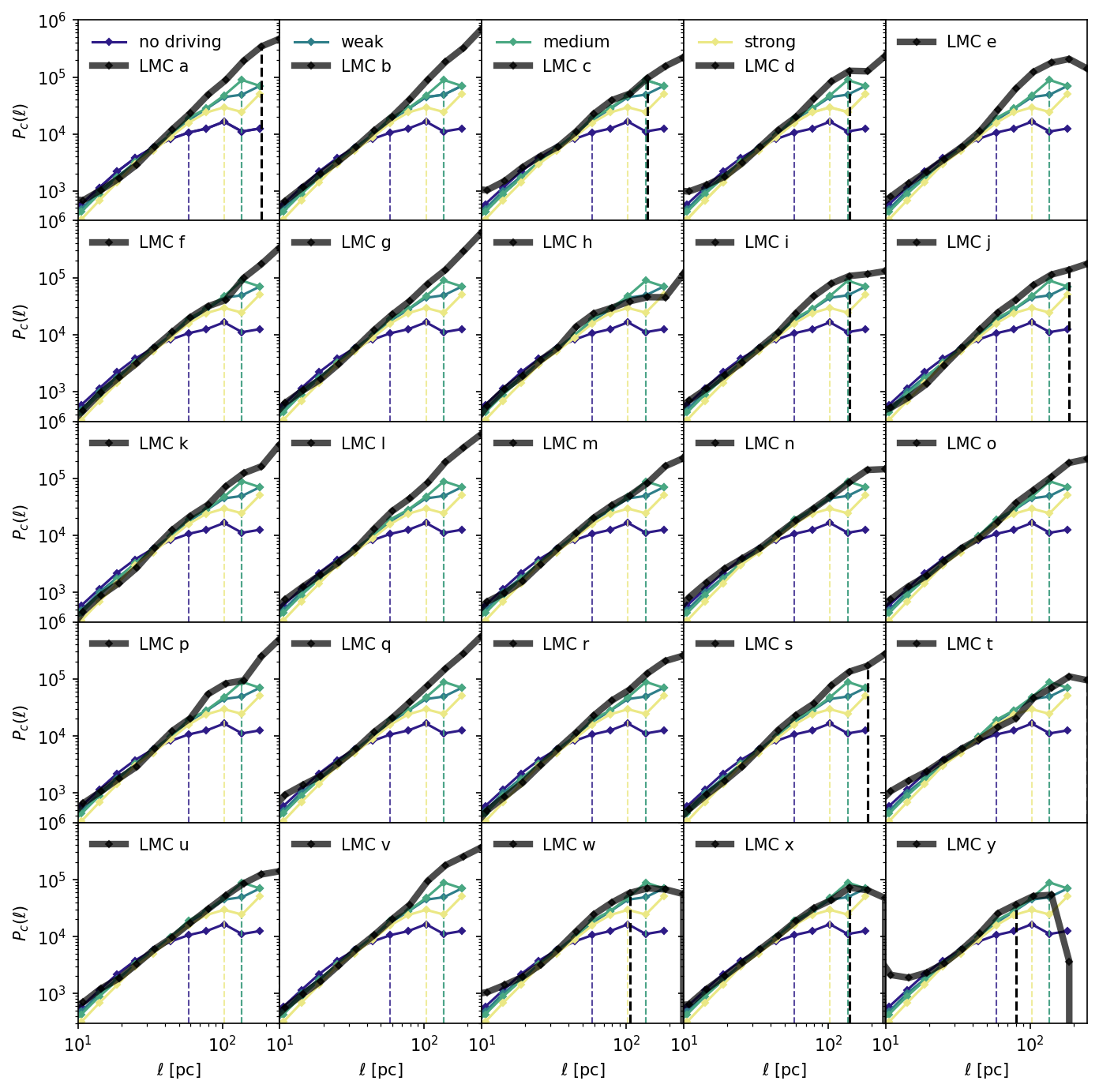}
    \caption{Comparison of the coherent power spectrum between simulations and individual regions in the LMC. The colored lines are the simulations for different driving strength. Each black line shows the coherent PS of the indicated LMC region scaled by an arbitrary factor to allow for easier comparison. The curved for the simulations are the same in every panel. The vertical lines mark the location of the turn-over scale $\ell_{\mathrm{LS}}$. If no vertical black line is shown, it means the turn-over in this part of the LMC occurs at scales larger than 250 pc.}
    \label{fig:compare_LMC_sims_PS}
\end{figure*}

Now that we know the signature of large-scale driving on the power spectrum decomposition, we can look for it in observations.
A comparison with the Milky Way is not straightforward because we practically see it edge-on, making it difficult to observe large scale structure.
The closest object for which suitable observations are available turns out to be the Large Magellanic Cloud (LMC).
Its orientation gives it an effective depth comparable to our 1 kpc simulation box and its proximity allow for a detailed investigation of the ISM structure.
We select the 500 $\mu$m observation by Herschel.
The image we use has been post-processed by \cite{Gordon_et_al2014} and made freely available to the community\footnote{\url{https://karllark.github.io/data_magclouds_dustmaps.html}}.
At a distance of 49.6 kpc \citep{2019Natur.567..200P}, the 14 arcsec observational resolution results in a physical resolution of 3.4 pc, comparable to the 
{\it physical} resolution 
reached for the dense gas in our simulations.  The latter can be broadly estimated to be 10 times the highest resolution which is about 0.24 pc.
At 500 $\mu$m, the observations trace thermal dust emission which is well-correlated to the underlying dust density distribution.
We assume the dust density to be proportional to the gas density through a constant dust-to-gas ratio.

\subsection{MnGseg applied on the LMC}
Before we run MnGseg on the full LMC, we cut the edges of the galaxy. The low signal-to-noise level in these regions makes the segmentation perform badly.
After this we are left with a 5 kpc $\times$ 5 kpc area.
Unlike the simulation, the real observation is not periodic.
To limit artifacts caused by the Fourrier transform, we add zero-padding of half the image size around the edges.
This is necessary to avoid leakage of bright sources at the edge of the image to the other side, which would create structures which are not present in the original image and produce a false excess of large scale power.
We then perform the MnGseg on this padded 25 kpc$^2$.
To match our simulation box, we reconstruct the power spectrum in several 1 kpc$^2$ regions indicated in Figure~\ref{fig:LMC_cuts}.
This can be done simply by averaging $\langle |\tilde{f}_{\ell, \theta}(\vec{x})|^2 \rangle_\theta$, which is outputted by MnGseg, over the corresponding region.
This is only possible due to the unique way in which the wavelet PS and its composition are calculated.
An additional advantage of this procedure is that it significantly improves the statistics on scales between 100 and 1000 pc.
Finally, we correct for the noise and Gaussian beam in the same way as in \cite{Robitaille_et_al2019}:
\begin{equation}
    P_\mathrm{measured}(k) = P_\mathrm{beam}(k) P_\mathrm{true}(k) + \mathrm{noise}
\end{equation}
\noindent where the noise is estimated by the total PS at the smallest scale considered.

Figure~\ref{fig:image_decomp_LMC} shows the decomposed images obtained with $q=2.0$.
The corresponding power spectra of each 1 kpc$^2$ region can be found in Figure~\ref{fig:LMC_PS_parts}.
The slope is measured by fitting the same range as for the large scale part of the simulation results.
On average we measure a total PS slope of -2.13 $\pm$ 0.30 with significant variations between regions.
This is in agreement with the recent study of \cite{Koch_et_al2020}, who report a global index of -2.18 $\pm$ 0.05 for the full LMC, with large local variations.
While the average value is roughly in agreement with the medium and strong driving simulations, the differences between regions indicate that a large variety of turbulence driving mechanisms might be at play in the LMC.

\subsection{Interpretation of the observed coherent PS}
Figure~\ref{fig:compare_LMC_sims_PS} compares the coherent PS of each LMC part with the results obtained from the simulations.
The LMC PS is scaled by an arbitrary factor for easier comparison.
This is necessary because 500 $\mu$m emission maps have different units than the column density maps of the simulation, and there is no straightforward conversion between the two quantities.
The first important observation to make is that \textit{none of the LMC regions matches the no-driving} simulation results.
This indicates that the observed coherent structures are larger than what can be generated by stellar feedback alone.
We see that for some regions the observations show a similar turn-over as for weak or medium driving case.
However, many regions show a continued increase beyond any turn-over seen in the simulation results.
This has several possible interpretations.
The first option is that the external driving experienced by the region is stronger than what we inputted in the simulations.
The fact that we do not see clear differences between the simulations with various driving strengths is an argument against this explanation.
A second option is that the injection scale is larger than what could be studied with the simulations.
Note that, while we injected turbulence on scales between 333 and 1000 pc, the statistics of the power spectrum limit our conclusions to scales below roughly 200 pc.
Since we do not vary the driving scales in our simulation suite, we cannot rule out or validate this possibility.
A third option is that we underestimated the statistical error for the PS analysis of the simulations and the turn-over we see between 100 and 200 pc for the simulations with driving is caused by low statistics rather than a true decline in coherent power.
This would suggest the coherent PS for the external driving simulations follows the same increasing trend as the regions in the LMC.
A last option is that the large scale structures in the LMC are not generated by large scale turbulence driving, but by a different physical process which was not included in the simulations.
The obvious culprit here could be variations in the galactic potential.
A way to test this could be to alter the background gravitational potential in our simulations to include such variations and rerun the simulation without driving.
Since this work focuses on large scales, we now make the assumption that the observed structure is indeed a consequence of large-scale driving but keep in mind that this may not be valid.

Overall, these results thus seem to indicate that there is significant large scale turbulence driving in all regions of the LMC.
The fact that the turn-over occurs on scales larger than what we reproduce in our simulations, seems to hint that the injection scale might be even larger than 1 kpc.
Possible large-scale driving sources in the LMC will be discussed below.

\subsection{Local variations of the turn-over scale}

\begin{figure}
    \centering
    \includegraphics[width=\columnwidth]{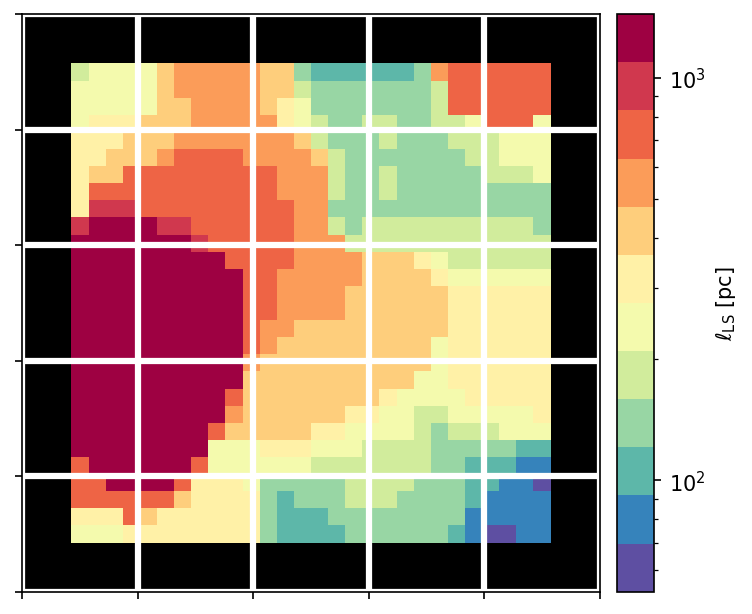}
    \caption{Map of the turn-over scale. Blue is consistent with no large-scale driving. Green shows regions consistent with values obtained from the simulations with external driving. Yellow corresponds to values around 250 pc, our statistics limit in the simulations. Orange and red indicate regions which have a turn-over scales larger than what could be measured in the simulations.}
    \label{fig:kLS}
\end{figure}

As we did for the simulations, we can measure the turn-over scale $\ell_{LS}$ by looking at which scales the coherent PS dominates over the Gaussian one.
Using the summation method to determine the PS in 1 kpc$^2$ regions, we are not limited to the 25 regions we defined in Figure~\ref{fig:LMC_cuts}.
We can select any region we want.
Pushing this philosophy to the limit, we sweep through the entire field of the LMC, shifting our 1 kpc$^2$ window each time by 44 pixels, resulting in 28 $\times$ 28 views spanning an area slightly less than 5 $\times$ 5 kpc$^2$.
In each view we reconstruct the PS and estimate the turn-over scale.
Figure~\ref{fig:kLS} shows the result.
Each point is centered in the 1 kpc$^2$ region we consider. Remark that points are correlated with their neighbors.
The edges have no data (black) because we do not consider views that go outside the central 5 kpc $\times$ 5 kpc area we selected.
Outside this area are the edges of the LMC and signal-to-noise ratios are poor.

Indicated with blue are the regions which have turn-over values consistent with no external turbulence driving.
Only the bottom right corner matches this regime.
The green regions found on the bottom, as well as a diagonal band stretching from region c-d towards i and j, have turn-over values similar to the ones obtained from the simulations with external driving (100 - 150 pc).
Yellow marks a value of 250 pc, the upper limit for the MnGseg analysis of the simulations.
Anything marked with orange or red has a coherent PS which increases beyond 300 pc.
Impossible to ignore is the giant red blob on the left side of the image which marks a region around 30 Doradus, the largest and most active star formation region in the LMC \citep{Harris_Zaritsky2009}.
We will come back to this in the next section.

This map clearly hints that all regions, except a small corner, contain structures larger than what can be generated by stellar feedback alone.
If we assume this large scale structure is indeed generated by large scale turbulence driving, then this map holds clues about which driving mechanisms operate in each part of the LMC.

\section{Discussion}
\label{sec:discussion}

\subsection{Possible sources of large-scale driving}
\label{sec:driving_sources}

The question arises: what are the primary sources of large-scale driving?
Here we speculate based on the results obtained in the previous section for the LMC.
Our goal is to give some qualitative first impressions.
More models are needed to be able to make solid conclusions.

\begin{figure}
    \centering
    \includegraphics[width=\columnwidth]{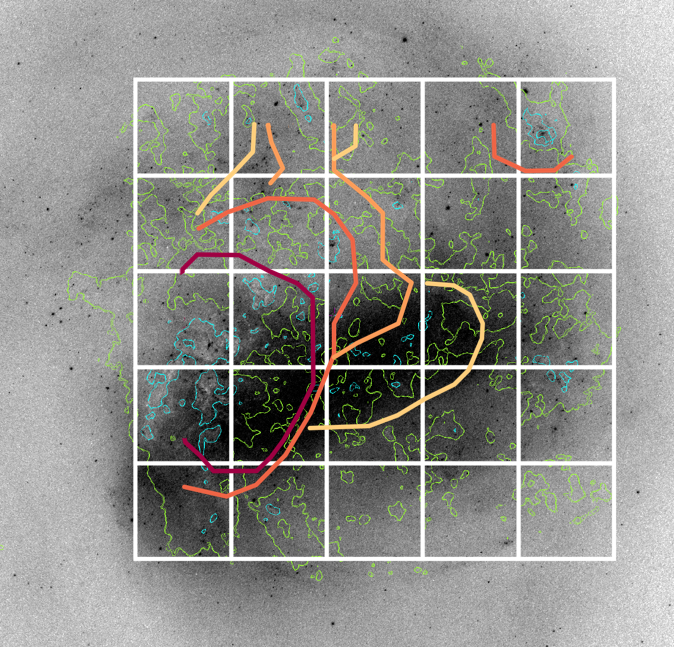}
    \caption{Stellar density map of the LMC from Gaia data release 2. The green and blue contours show the 500 $\mu$m dust emission. The orange lines mark the regions with large turn-over scale. The grid indicates the location of our 1 kpc squares.}
    \label{fig:Gaia_LMC}
\end{figure}

\begin{figure}
    \centering
    \includegraphics[width=\columnwidth]{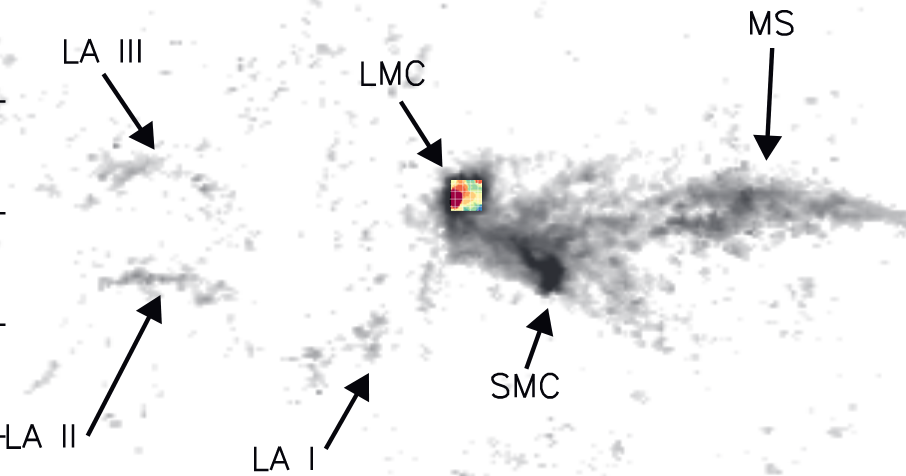}
    \caption{HI column density map from \citet{Nidever_et_al2008}
    showing the interaction with the Small Magellanic Cloud, and the different parts of the Leading Arm which is due to the interaction with the Milky Way. We insert Figure~\ref{fig:kLS} at the appropriate position.}
    \label{fig:environment_LMC}
\end{figure}

It might not seem clear at first what could be the source of large scale turbulence energy injection in an irregular galaxy like the LMC.
However, the stellar component actually contains a bar spanning a large part of the galaxy and one spiral arm which can be seen in Figure~\ref{fig:Gaia_LMC} \citep{Harris_Zaritsky2009, Helmi_et_gaia2018}.
This galactic stellar dynamic affects the gas through changes in the gravitational potential.
It is possible that turbulence taps energy from this large scale gravitational reservoir.
The bar spans almost the entire width of the LMC and is located slightly below the center of our analysis domain.
Interesting observations can be made when comparing Figure~\ref{fig:kLS} to the stellar density image by Gaia (Figure~\ref{fig:Gaia_LMC}).
To make the comparison easier, we draw orange contours which mark the main features of the turn-over scale map.
The light orange diagonal patch in the center matches well with the location of the stellar bar.
Whereas any large-scale driving signature in the top corners of the image could be matched with the spiral arm, the correspondence is not as clear as for the bar.

The LMC is also tidally interacting with the Small Magellanic cloud (SMC) and the Milky Way \citep{donghia16, petersen21, Lucchini_et_al2021}.
This can possibly inject energy on the full scale of the galaxy with some preferential direction.
In Figure~\ref{fig:environment_LMC} we show the large scale environment of the LMC.
The SMC is connected to the LMC through the Magellanic bridge, visible in HI.
The other prominent features are the Magellanic Stream (MS), marking the trail of the orbit of the LMC and SMC around the Milky way, and the Leading Arm (LA).
These structures are thought to be the result of the interaction between the LMC and SMC and with the Milky Way \citep{donghia16}.
\citet{Nidever_et_al2008} find that some parts of the MS and LA can be traced back to an HI overdensity in the southeast part of the LMC, where also 30 Doradus is located.
This coincides with the location of the big red blob in Figure~\ref{fig:kLS} where, in our interpretation, we find a signal of turbulence driving with a very large injection scale.
One interpretation of the link between the MS/LS and the starburst region 30 Doradus is that some parts of the MS and LS are created by gas outflow from extreme stellar feedback \citep{Nidever_et_al2008}.
Our results show a clear correlation between large scale driving and 30 Doradus.
This can be interpreted in two (possibly more) ways: the large scale driving mechanism generating the signal in Figure~\ref{fig:kLS} is
\begin{itemize}
    \item[1)] caused by 30 Doradus itself,
    \item[2)] caused by another mechanism which possibly also created 30 Doradus, whose origin is still debated.
\end{itemize}

\begin{figure}
    \centering
    \includegraphics[width=\columnwidth]{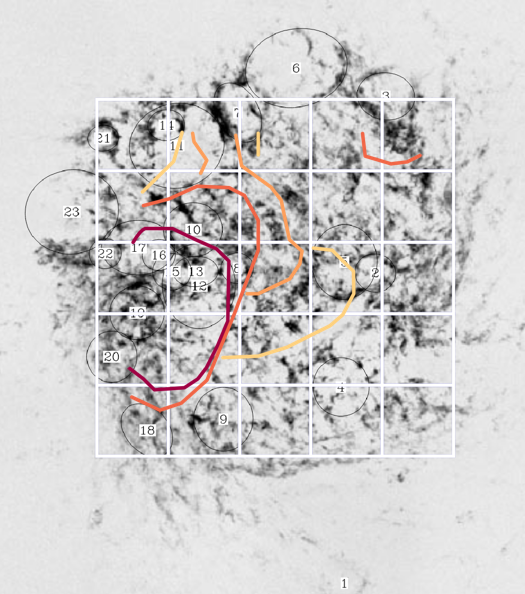}
    \caption{Approximate overlay of our selected LMC regions with the HI map and identified super giantshells from \citet{Kim_et_al1999}. The orange lines mark the regions with large turn-over scale.}
    \label{fig:superbubbles}
\end{figure}

This brings us to the last possibility, which is the existence of supergiant shells.
HII bubbles from individual stars are typically of the order of a few tens of parsec.
Individual supernova bubbles are larger with total sizes of up to 100 pc.
Supergiant shell are created when multiple supernova go off in close proximity of one another, combining several bubbles to generate structures of the order
of several hundreds of pc and up to 1 kpc \citep{maclow1989,Chu2008}.
This typically requires multiple generations of star formation.
It is unclear exactly how common these are.
Several super shells have been identified in the LMC in HI emission maps.
In Figure~\ref{fig:superbubbles}, we show the data from \cite{Kim_et_al1999} overlaid by our 1 kpc regions.
We inspect each region and look for a link between the large-scale driving signature and the presence of a supergiant shell.
The left side of the image is dominated by several shells.
Indeed, 30 Doradus has been identified as the source of large ionized bubbles.
Any red or orange region in sections a, b, f, g, k, l and p could be explained by a supergiant shell.
Regions more in the center of the galaxy (third column), do not contain any known supergiant shells.
They do however show strong signs of large scale turbulence driving.
On the other hand, regions n, x, and s have bubbles but no clear signature of them in the $\ell_{\mathrm{LS}}$ map.
Overall, we see that on the left side of the LMC, supergiant shells could potentially be the origin of the large scale driving.
However, in the central and right side there is no clear correlation between the location of supergiant shells and large scale driving.

These observations seem to indicate that supergiant shells alone cannot explain all of the large scale coherent structure we see.
Note that in principle these types of shells can form in our simulations, since we do include supernova feedback in a self-consistent manner.
In particular supernovae explode in a correlated way since they are associated with the dense molecular clouds.
So the no-driving simulation should include the signature of these super giant shells.
Intuitively, we could expect the size distribution for shells to leave an imprint on the coherent PS. 
If there was a sharp cutt-off at a certain maximum size, we would not detect any power in the coherent PS above this scale for the no-driving simulations.
The fact that we see a flattening rather than a cut-off suggests the simulations do contain several shells which are larger than the typical SN radius.
There are however a few possible caveats that we discuss below.

\subsection{Caveats related to numerical algorithms}
As usual, there are many caveats related to the simulations.
Our simulations have only one driving spectrum, which peaks at scales of half the box size.
It is to be expected that different driving modes will create different structure sizes.
The exploration of different driving modes and scales is left to future work.

Individual supernovae are not resolved in the simulation.
We thus have to rely on a subgrid recipe.
While extensively tested in previous simulations, a subgrid model is usually at best a reasonable approximation.
It is for instance possible that our recipe is not able to reproduce a realistic number of supergiant shells, in which case it is possible that the source of the large-scale driving due to stellar feedback could be underestimated.
For example, it is known that the exact location at which supernovae explode, plays an important role in how efficiently it can inject energy and shape the ISM \citep[e.g.][]{Hennebelle_Iffrig2014,gatto2015}.
One possibility is that our recipe
to estimate how far massive stars can travel from their parent cluster, leads to less efficient giant shell creation.
Another possibility is that a volume of 1 kpc$^3$ is not enough to allow for supergiant shells to occur.
Finally, our recipe only injects momentum, not energy directly, in contrast to the recipe used for 
instance in \citet{kim2017}. 

We also do not include feedback from stellar winds, because this is computationally expensive.
While this might have an impact on small scales
\citep[see however][who argued that winds may actually reduce the extension of HII regions]{geen2022}, we expect the impact on large scales to be minimal.

\subsection{Caveats related to the interpretation}

The simulations tell us that large-scale driving generates large scale structure.
However, this does not necessarily imply that an observed large scale structure is generated by a large scale turbulence driving mechanism.
The structure could have been generated by a process which is not present in the simulations and which leave a similar signature on the spatial power spectrum but does not inject additional energy to drive the turbulence.
Galactic potential variations could be a candidate, though it is unclear whether or not these contribute to turbulence driving.

Using MnGseg, we can determine that there is structure of a certain size present, but we have no direct information on which process was responsible for creating it.
We need to indirectly infer this by using the results from our simulations.
By modeling regions with different galactic potentials, different driving scales and other variations in driving parameters, we might find more clues to explain the observations and put more constrains on the origin of structure and turbulence in the ISM.
Future work which studies the turbulence cascade in full galaxy simulations will also shed more light on this matter.

Another issue is that the properties of the density field give only an indirect measure of the turbulence properties. 
To gain a more complete picture, the velocity field could be analysed. This is however beyond the scope of this work.

Lastly, in purely 2D turbulence an inverse cascade occurs due to conservation of vorticity \citep{Kraichnan1967, Lazarian_Pogosyan2000}.
This results in an energy PS which becomes steeper than the Kolmogorov PS for scales larger than the injection scale.
This could complicate the interpretation of our results since we interpret changes of the coherent PS as a signature of the injection scale (assuming the density PS reflects the energy PS).
We do not know how this inverse cascade would affect the decomposition of the PS into coherent and Gaussian part.
However, the galactic disk is not a perfectly 2D structure; it has a thickness.
Also in the ISM there are other sources of angular momentum besides the turbulence injection mechanisms.
The presence of gravity and magnetic fields complicates the situation.
It is unclear whether the inverse cascade can still happen under these conditions.
Indeed, \citet{Hennebelle_Audit2007} verified using 2D simulations that when the fluid is non-barotropic, as is the case in the large-scale ISM, there was no signature of an inverse cascade.
On the otherhand, an identical but isothermal simulation gave an energy PS much closer to the prediction from \citep{Kraichnan1967}.
In our simulations, the 2D external driving has injection scales which are already comparable to the box size.
This leaves little room for an inverse cascade.
The stellar feedback bubbles usually remain smaller than the disk thickness and thus are a 3D source of driving.
We also do not observe a steepening of the PS at scales larger in Figure~\ref{fig:powerspectra_decomposition}, though it is possible that the effect would only be visible on scales larger than what we can analyse with our setup.
On the observational side, \citet{Koch_et_al2020} show that the spatial PS of the LMC, as well as other galaxies in their sample, can be described by a single power-law without breaks.
All this suggest that the inverse cascade is not important for the interpretation of our results.

\subsection{Comparison to the literature}

Multiple authors have studied the origin of turbulent velocities in the ISM.
Using the velocity coordinate spectrum (VCS) technique, \citet{Chepurnov_et_al2010} determined the turbulence properties of a high-latitude region in the Milky Way.
This technique analyses the velocity information in the HI line emission data.
They derive an injection scale of 140 $\pm$ 80 pc.
The big uncertainty on the result illustrates how difficult it is to study large scale processes in the Milky Way.
\citet{Stanimirovic_Lazarian2001} investigated the turbulence in the SMC, but could not identify an injection scale. They concluded that it is likely of the size of the galaxy or even larger. Later, using again the VCS technique, \citet{Chepurnov2015} indeed recovered a very large injection scale of 2.3 kpc which implies the presence of large scale turbulence driving, possibly due to tidal interaction with the LMC or supergiant shells.
Another example can be found in a study by \citet{Dib_Burkert2005}, where they find an injection scale of about 6 kpc for the irregular galaxy Holmberg II.
These findings are qualitatively in line with our analysis of the LMC, where some regions show large values of $\ell_\mathrm{LS}$. It is not straightforward to compare these results quantitatively, since we did not vary the injection scale in our simulations and thus cannot tell for certain how $\ell_\mathrm{LS}$ depends on this.

Several studies also looked for the origin of ISM turbulence in M33.
\citet{Utomo_et_al2019} obtained an estimate of the turbulent energy density of roughly 1 - 3 $\times 10^{52}$ erg pc$^{-2}$ which is of the same order of magnitude as our estimate for the energy injected by SN in our simulations, if we assume a low efficiency.
In their work, they concluded that SN are indeed the main source of turbulence within 8 kpc.
In the outer regions of the galaxy, magneto-rotational instability could provide the additional energy observed.
However, \citet{Koch_et_2018} obtain a different picture, where even the combination of the two cannot explain the outer parts.
Another possible source of turbulence in M33 would be due to the interaction with M31. \citet{Utomo_et_al2019} could not exclude accretion as a source when they applied the model of \citet{klessen10}.
However the adopted parameters have large uncertainties.
Interesting to note is that M33 has no bar. If bars are the main large-scale turbulence driving mechanism in the inner parts of a galaxy, we would indeed expect an energy density compatible with stellar feedback as main driving mechanism for bar-less galaxies.

In a sample of dwarf galaxies, \citet{Stilp_et_al2013} found that star formation feedback alone is not enough to explain the observed HI kinematics and that another source of turbulence is needed to explain their data.
Previous studies thus already showed that a single dominant source of turbulence is not enough to explain the observed turbulent velocities across all environments.
The results in this study further support this.

\section{Conclusions}
In this paper we have performed a series of four simulations describing a stratified star-forming interstellar medium modelling a 1 kpc region of a galaxy.
An external large scale turbulence driving is applied to mimic the possible influence of large galactic scale energy injection, which we cannot self-consistently represented in our chosen computational domain. 
Using a multi-scale non-Gaussian segmentation technique called MnGseg on the column density maps of our simulations, we have computed the power spectrum of the  coherent structures and the Gaussian background that MnGseg allows to separate and identify.
We found that the power spectrum of the coherent structures carries a signature of the presence of external large scale driving.
In the case where the turbulence is driven by stellar feedback only, the coherent power spectrum is a powerlaw which flattens above a turn-over scale of 60 pc, the typical radius of a single supernova.
When external large scale driving was included, the turn-over scale shifted by a factor $\approx 2$ to larger scales.
In our simulations, external large scale turbulence driving was thus able to create more large scale structure than what can be created by stellar feedback alone.

We applied the same technique on 500 $\mu$m observations of the LMC, which we divided in several 1 kpc$^2$ regions, and found that only 1 out of 25 regions matched the results from the no-driving simulation.
This result seemingly indicates that some form of large scale turbulence driving is present in almost all parts of the LMC.
To obtain clues about the nature of these driving mechanisms, we studied local various of the turn-over scale in detail.
We found a particularly large turn-over scale around the starburst region 30 Doradus.
This signal could be explained by a collection of supergiant shells, a form of extreme stellar feedback not observed in the simulations.
Supergiant shells are also present in other regions of the LMC, but there they did not correlate with a large scale driving signal.
Also, many regions of the LMC do not contain supergiant shells and we still observed signs of large scale driving in here.
We did see a potential correlation with the stellar bar, hinting at the possible importance of galactic stellar dynamics as a turbulence driving mechanism.

The results in this work lead us to conclude that regular stellar feedback is not enough to explain the observed ISM structure on scales larger than 60 pc.
Extreme feedback in the form of supergiant shells likely plays an important role but cannot explain the results in all the regions of the LMC.
If we assume the ISM structure is generated by turbulence, another large scale driving mechanism is needed to explain the entirety of the observations.

\section*{Acknowledgements}
This work was granted access to HPC
   resources of CINES and CCRT under the allocation x2020047023 made by GENCI (Grand
   Equipement National de Calcul Intensif).
   This research has received funding from the European Research Council
synergy grant ECOGAL (Grant  855130).
TC would like to thank all the members of the ECOGAL collaboration for discussions and comments which greatly improved this work.
This work also benefited from the Grand Cascade workshop hosted at Institute Pascal in Orsay, France.
TC would also like to thank Steve Longmore for teaching her how to use Aladin during the Advanced School on Star formation in Granada.
RSK and SCOG acknowledge funding from the German Research Foundation (DFG) under Germany’s Excellence Strategy for the STRUCTURES cluster of excellence in Heidelberg (EXC-2181/1 - 390900948) and also via the collaborative research center (SFB 881, Project-ID 138713538) ``The Milky Way System'' (subprojects A1, B1, B2, and B8). LT acknowledges funding from the Italian Ministero dell'Istruzione, Universit\`a e Ricerca through the grant Progetti Premiali 2012 – iALMA (CUP C$52$I$13000140001$), from the Deutsche Forschungs-gemeinschaft (DFG, German Research Foundation) - Ref no. 325594231 FOR $2634$/$1$ TE $1024$/$1$-$1$, from the DFG ORIGINS cluster of excellence  in Munich (www.origins-cluster.de), and from he European Union's Horizon 2020 research and innovation programme under the Marie Sklodowska-Curie grant agreement No 823823 (DUSTBUSTERS).

\section*{Data Availability}
The data underlying this article are available in the Galactica database at \url{http://www.galactica-simulations.eu/db/}, and can be accessed with the tag \texttt{LS\_DRIVING}.
All analysis scripts used to produce the figures in this work are available at \url{https://bitbucket.org/TineColman/apophis/}.
in the directory \texttt{Paper\_Colman\_et\_al2022}.


\bibliographystyle{mnras}
\bibliography{large_scale}



\newpage

\appendix

\section{How MnGseg works}
\label{sec:mngseg_appendix}

\subsection{Wavelet power spectrum}

\begin{figure*}
    \centering
    \includegraphics[scale=0.75]{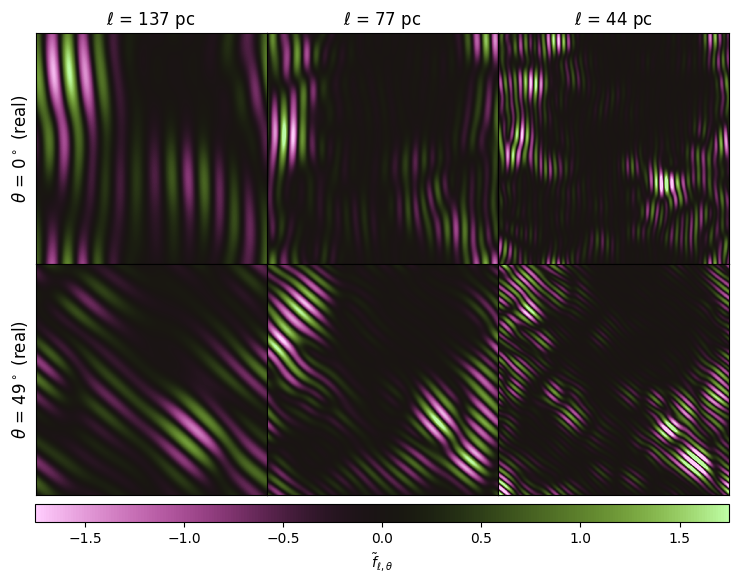}\\
    \includegraphics[scale=0.75]{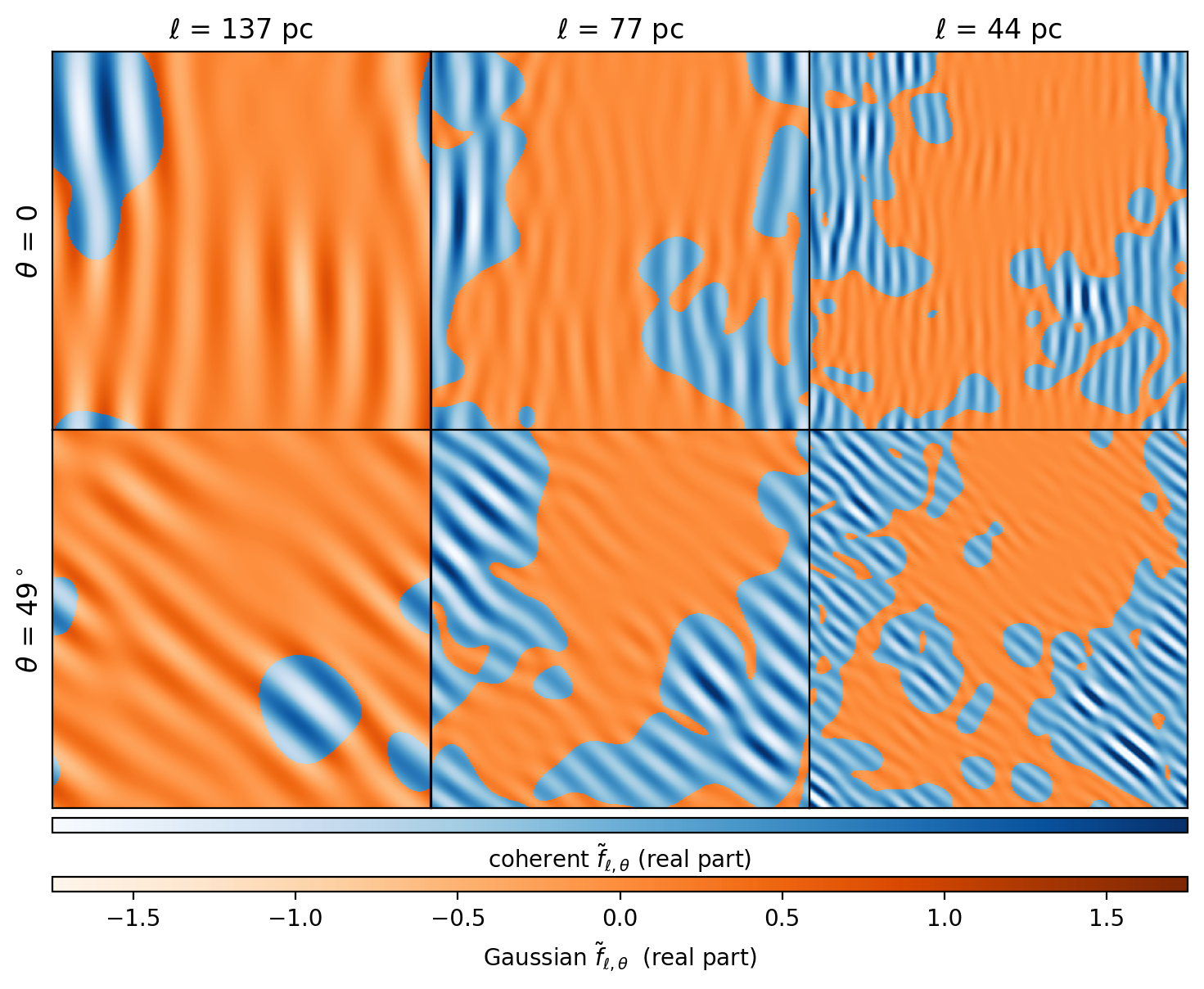}
    \caption{Top: Complex wavelet coefficients obtained after convolving the original image (left panel of Figure \ref{fig:how_mnGseg_image_reconstruction}) with a Morlet wavelet of various sizes and angles (see Equation~\ref{eq:wavelet_coeff}).
    Bottom: top panel divided into its Gaussian and coherent parts.}
    \label{fig:how_mnGseg_Morlet}
\end{figure*}

\begin{figure*}
    \centering
    \includegraphics[scale=0.85]{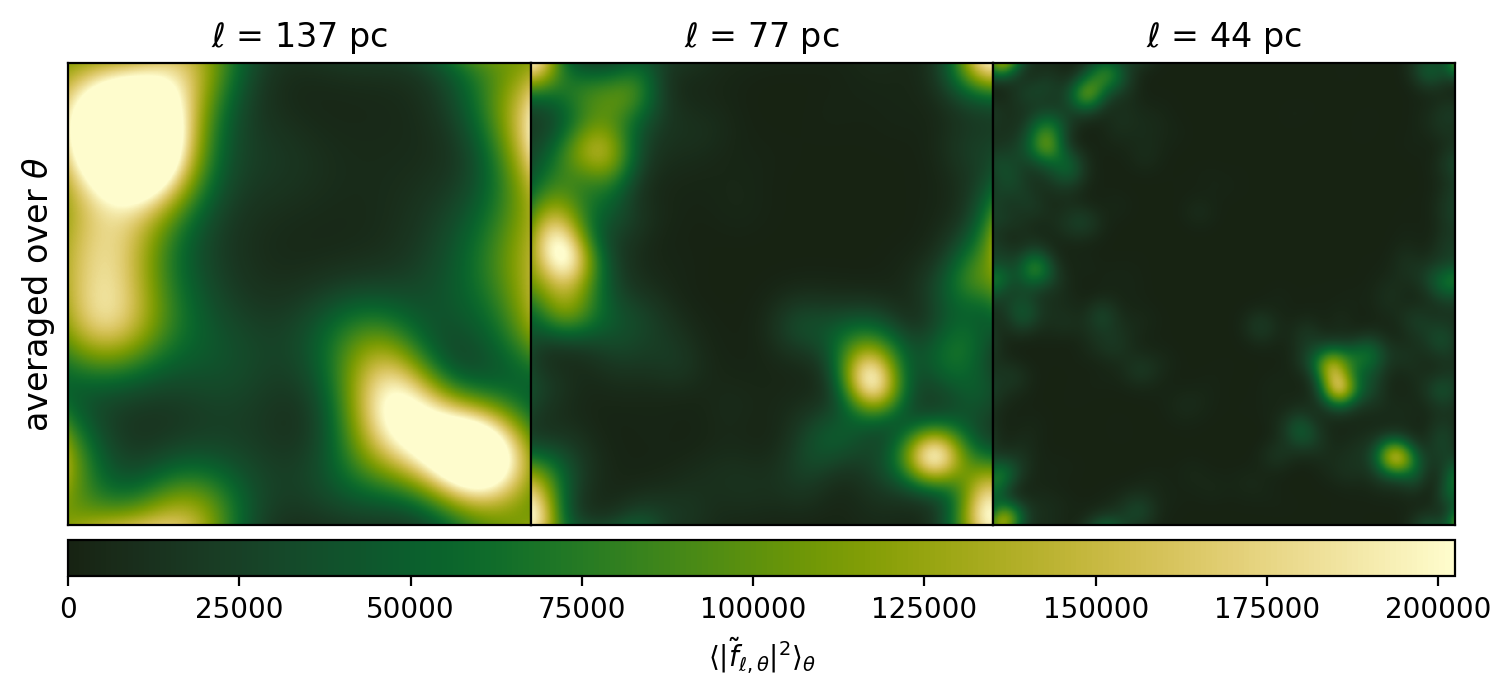}\\
    \includegraphics[scale=0.95]{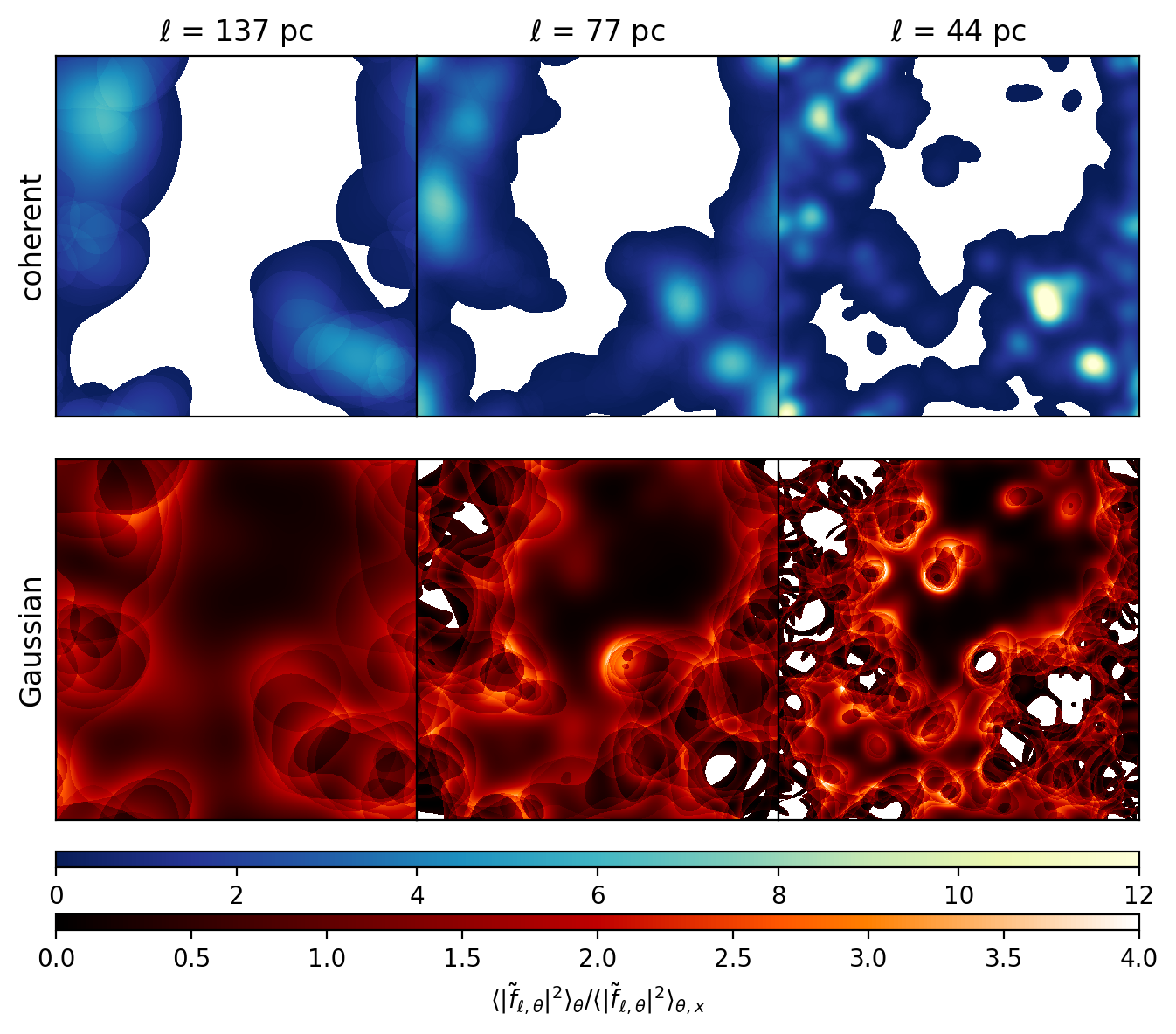}
    \caption{Top: Square of the modulus of the wavelet coefficients for the scales shown in Figure~\ref{fig:how_mnGseg_Morlet} after averaging over all angles. The $\pi/2$ phase shift between its real and its imaginary parts eliminates the wavelet's oscillations. We see structures of various scales are recovered at the locations which were highlighted in the previous figure.
    Bottom: The equivalent of the top panel after segmentation.}
    \label{fig:how_mnGseg_average_angles}
\end{figure*}

\begin{figure*}
    \centering
    \includegraphics[scale=0.59,trim={0 -1cm 0 0},clip]{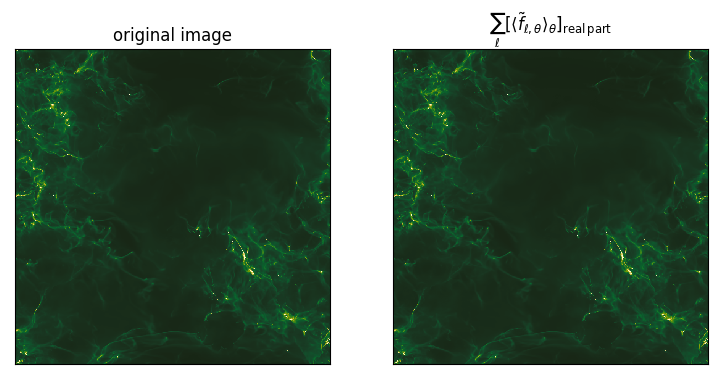}
    \includegraphics[scale=0.6]{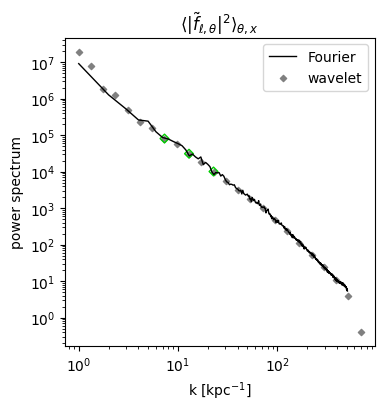}\\
    \includegraphics[scale=0.59,trim={0 -1cm 0 0},clip]{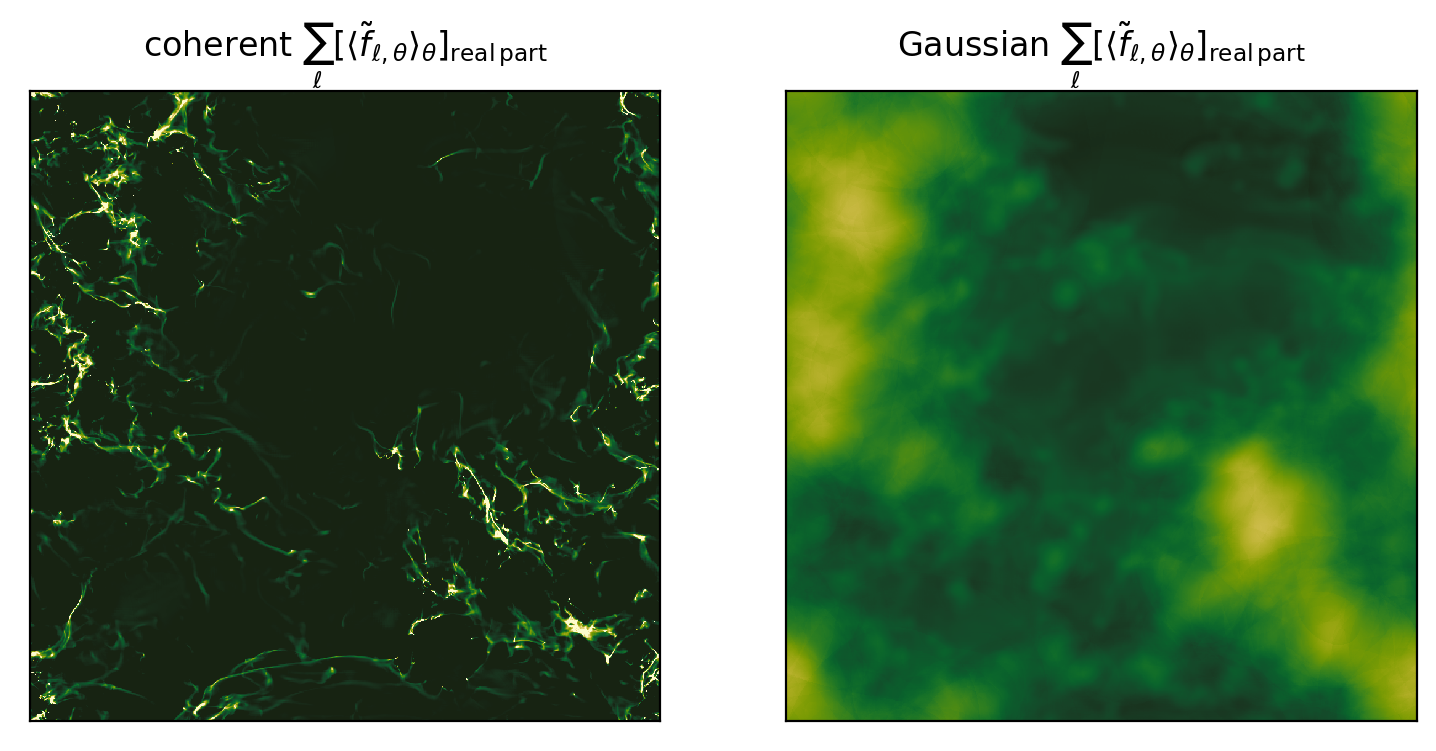}
    \includegraphics[scale=0.6]{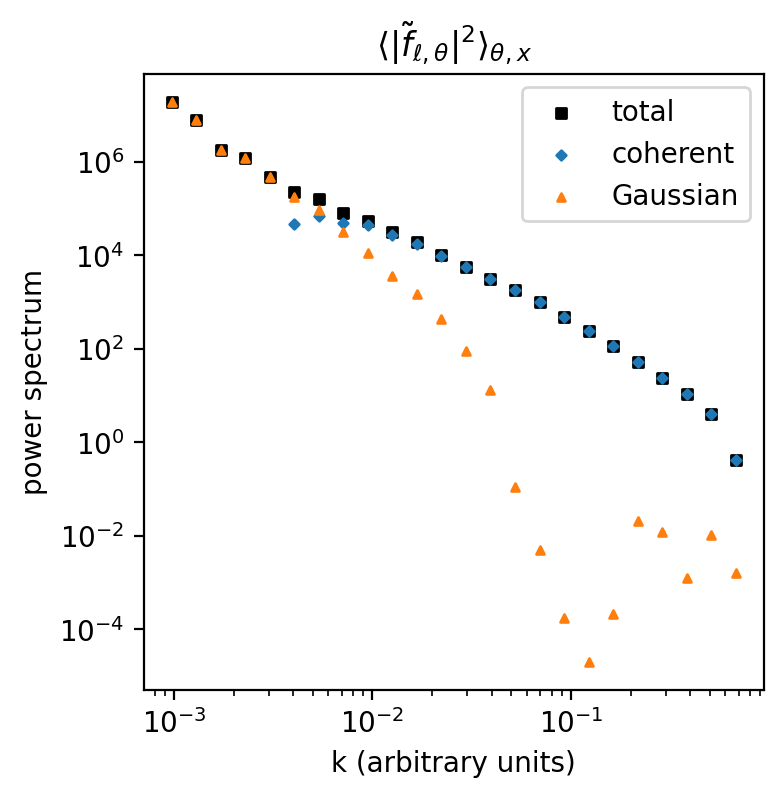}
    \caption{The power spectrum (top right panel) can be calculated by averaging the squared wavelet coefficients over angle and space (see Eq~\ref{eq:wavelet_powspec}). Note that this is equal to averaging over space in Figure~\ref{fig:how_mnGseg_average_angles} (top); we marked the corresponding scales in green.
    As illustrated in the top middle panel, by summing up the contributions from each scale (real-valued wavelet coefficients only), we can recover the original image which is shown on the top left.
    The bottom row shows the separated image of the coherent and Gaussian component. Here, the Gaussian image contains the average of the original image, so that coherent image + Gaussian image = original image. The bottom right panel shows the segmented power spectra.}
    \label{fig:how_mnGseg_image_reconstruction}
\end{figure*}

\begin{figure*}
    \centering
    \includegraphics[width=\textwidth]{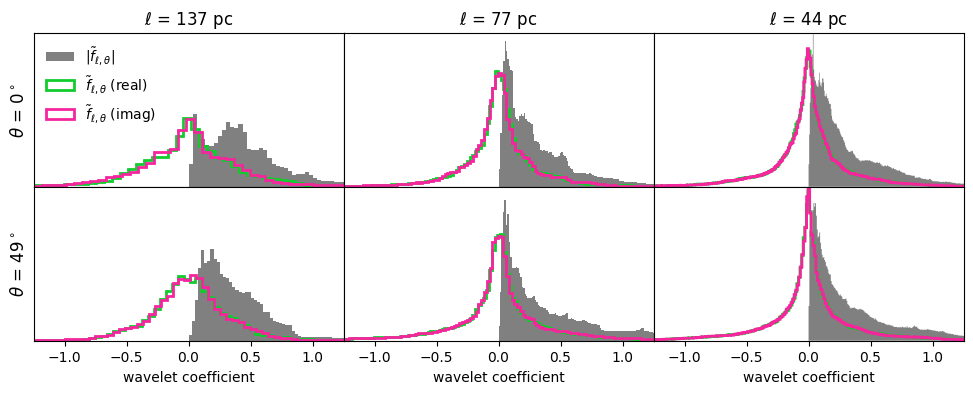}
    \caption{Histograms of the real and imaginary part of the wavelet coefficients from the images in Figure~\ref{fig:how_mnGseg_Morlet}. For a pure fractal both the real and the imaginary part have a Gaussian distribution, while the modulus is Rician. The presence of coherent structures alters the distributions.}
    \label{fig:how_mnGseg_histogram}
\end{figure*}

\begin{figure*}
    \centering
    \includegraphics[width=\textwidth]{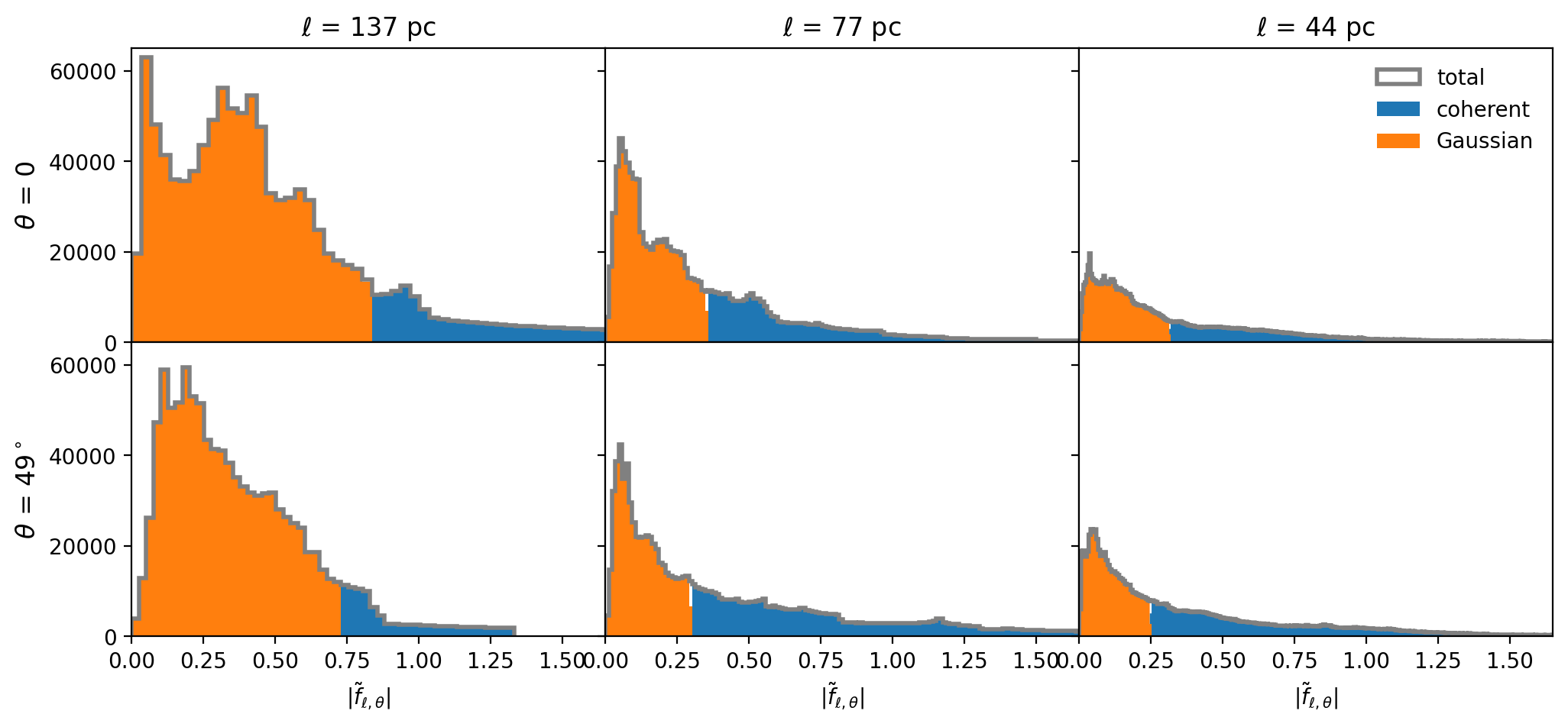}
    \caption{Segmentation into Gaussian and coherent parts.}
    \label{fig:how_mnGseg_segmentation_histogram}
\end{figure*}

Since MnGseg has not been widely applied yet, in particular to numerical simulations, we provide
a series of figures to illustrate in details the 
various steps of the technique.

\begin{itemize}
    \item Step 1: Convolution with Morlet wavelet for different sizes and angles. The result is illustrated in Figure~\ref{fig:how_mnGseg_Morlet} (top). What is displayed is the real part of the complex wavelet coefficients.
    \item Step 2: Calculate the square of the coefficients.
    \item Step 3: Average over angles (Equation \ref{eq:wavelet_powspec}). The result for several scales is shown in Figure~\ref{fig:how_mnGseg_average_angles} (top).
    \item Step 4: To calculate the wavelet power spectrum, we now simply have to average over space. The result is shown in Figure~\ref{fig:how_mnGseg_image_reconstruction} on the top right. We see that the wavelet power spectrum is equivalent to the classical Fourier power spectrum. Because the original image is real-valued, one can also reconstruct the original image by summing only the real part of the wavelet coefficients over all scales. This is shown in the top left and middle panel of Figure~\ref{fig:how_mnGseg_image_reconstruction}.
\end{itemize}

\subsection{Decomposition into Gaussian and coherent part}

When taking the histogram of the convolved images, both the real and imaginary part of the wavelet coefficients are distributed Gaussian for a fractal.
Consequently, the modulus is a Rician distribution.
This is illustrated in Figure~\ref{fig:how_mnGseg_histogram}.
The segmentation is done on the modulus $|\tilde{f}_{\ell,\theta}(\vec{x})|$.
Using a parameter value $q$ (see Section~\ref{sec_MnGseg_theory}), the distribution is split into two parts, the bulk being counted as Gaussian, while the tail is assigned to the coherent part.
This is illustrated in Figure~\ref{fig:how_mnGseg_segmentation_histogram}.
Figure~\ref{fig:how_mnGseg_Morlet} (bottom) shows the corresponding segmentation in the real part of the convolved image.
After averaging over angles, we obtain the bottom part of Figure~\ref{fig:how_mnGseg_average_angles}.
The coherent part of this figure makes it clear why this technique works well for core extraction, as the Gaussian noise is removed and a clean signal of an overdensity of a certain size can be obtained.
For large scales we see the image is not completely smooth and some artifacts remain in the coherent part.
This is due to the limited statistics on the largest scales.

Again we can reconstruct the total image by summing the real parts over all scales, but now we have two separate images: one for the coherent part and one for the Gaussian part (Figure~\ref{fig:how_mnGseg_image_reconstruction}, top left and middle).
The corresponding power spectra are shown in the top right panel of Figure~\ref{fig:how_mnGseg_image_reconstruction}.

\section{Varying the q parameter for segmentation}
\label{sec:test_q}

\begin{figure*}
    \centering
    \includegraphics[scale=0.48]{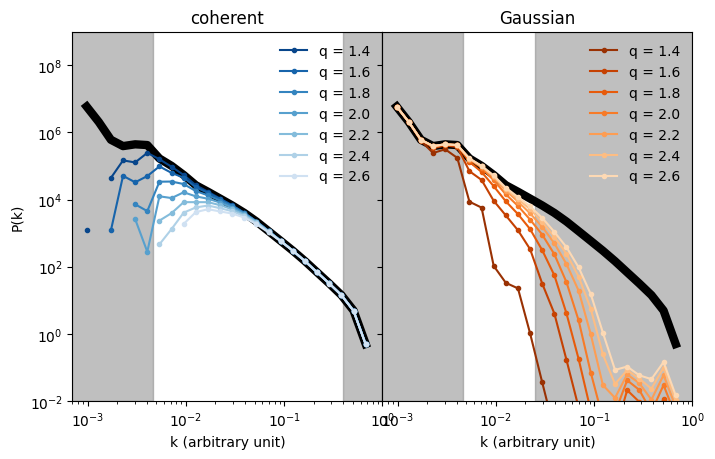}
    \includegraphics[scale=0.48]{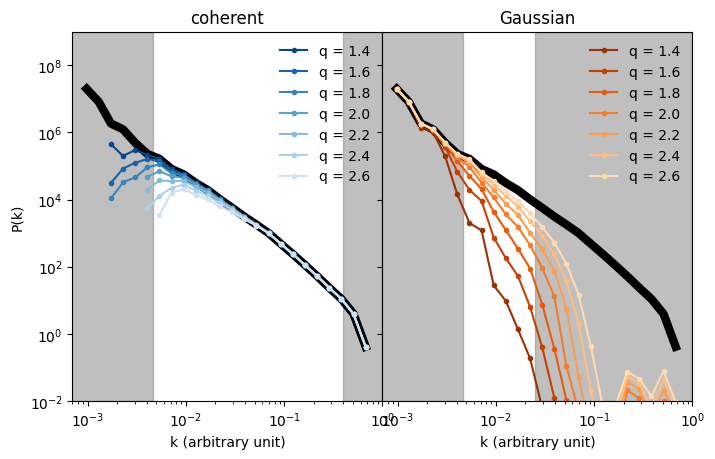}\\
    \includegraphics[scale=0.48]{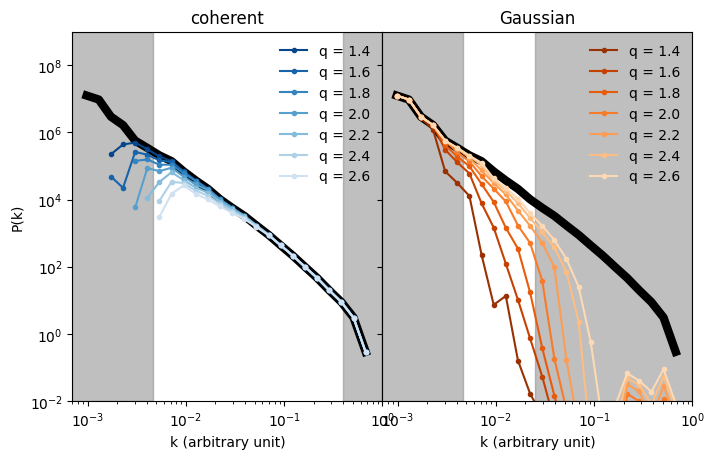}
    \includegraphics[scale=0.48]{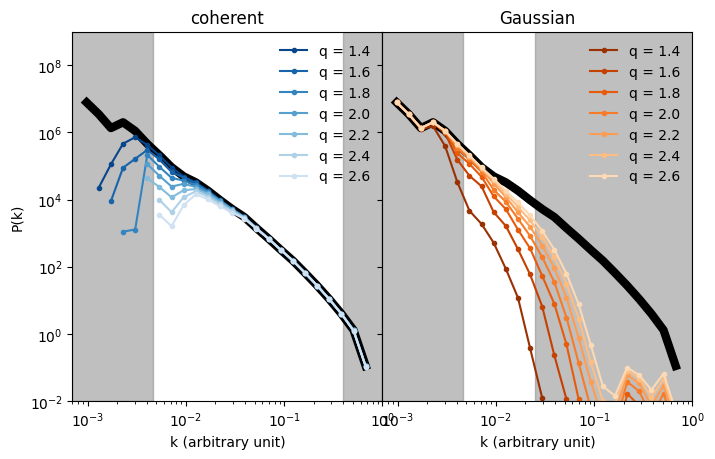} 
    \caption{Varying $q$ for simulations. Top left: no driving, Top right: weak driving, bottom left: medium driving, bottom right: strong driving. The black line is the total PS. The grey regions indicate the statistics limits with an error of maximum 20\% (left) and the resolution limits (right). The resolution limit for coherent structures is equal to the 10 times the finest resolution of the grid, whereas for the Gaussian it is 10 times the coarsest grid resolution.}
    \label{fig:fixed_q}
\end{figure*}

As discussed in Sect.~\ref{sec_MnGseg_theory}, the separation between a Gaussian and a coherent 
part relies on the free $q$ parameter defined in 
Eq.~(\ref{eq:q_parameter}) whose value has to be 
selected. Here we illustrate how the results 
depend on its value.

In Figure~\ref{fig:fixed_q} we show the resulting segmentation for our four simulations using different values for $q$.
We see the segmentation depends on $q$ but in a systematic way.
Comparing results obtained with the same $q$ is thus a reasonable thing to do.
When $q$ is low, the definition of Gaussian is very strict and as a result almost everything is considered to be coherent structure.
On the other hand, when $q$ is large, Gaussianity is only loosely defined and the coherent component is thus small.
Alternatively one can say that in this case the definition of a coherent structure is very strict.
Indeed, it is difficult to recover large scale coherent structures with large $q$.
It is important to note is that there is a clear difference between the behaviour for the no-driving simulation and the simulations with external turbulence driving.
Even with a very large $q$, the simulations with external driving produce a strong coherent signal in the power spectrum at large scales close to the statistics limit.
For the no-driving case, however, the coherent PS flattens already at smaller scales.

\section{Padding and apodizing}
\begin{figure*}
    \centering
    \includegraphics[width=\textwidth]{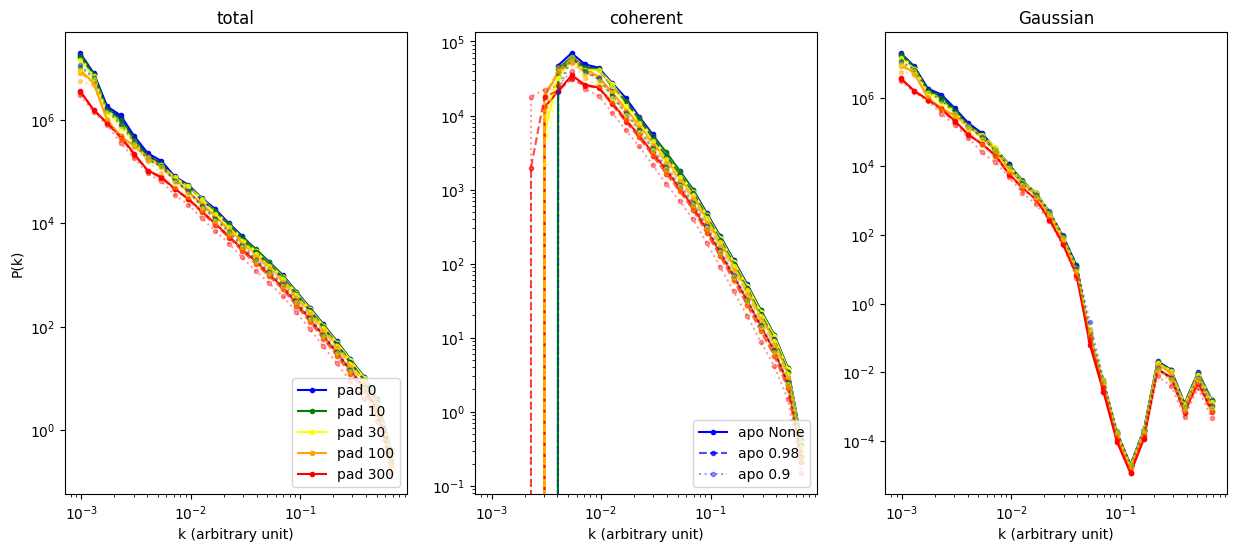}
    \caption{Testing various values for padding and apodizing on the simulation image (weak driving, $q=2$).}
    \label{fig:pad_sim}
\end{figure*}

When applying MnGseg to the LMC data a difficulty 
arises because unlike for the numerical simulations, it is not periodic.
To estimate how much the results depend on zero-padding and apodizing which are typically used to avoid artifacts created by Fourier operations on non-periodic data, we apply these techniques to the simulation data and rerun the MnGseg analysis.
Figure~\ref{fig:pad_sim} portrays the resulting 
PS and its decomposition using 
several padding and apodizing values. 
As can be seen the results weakly depend on
the specific choice that is being made.
The slope of the PS tends to be steeper when no padding is applied.
This is likely due to spurious large scale structure which is introduced when a bright signal at the edge of the image leaks to the other side.
For this reason, we decide to use a large padding when analysing the LMC.
If we choose no padding, or the padding applied is too small, 30 Doradus appears also on the right hand side of the image.

Apodizing the edges of the LMC image does not seem to affect the power spectrum. However, experimentation shows that it nevertheless introduces spurious structure at the borders of the image.
For this reason, we decide to not apply any apodization.

\section{Statistical stationarity}
\label{sec:appendix_steady_state}

\begin{figure}
    \centering
    \includegraphics[width=\columnwidth]{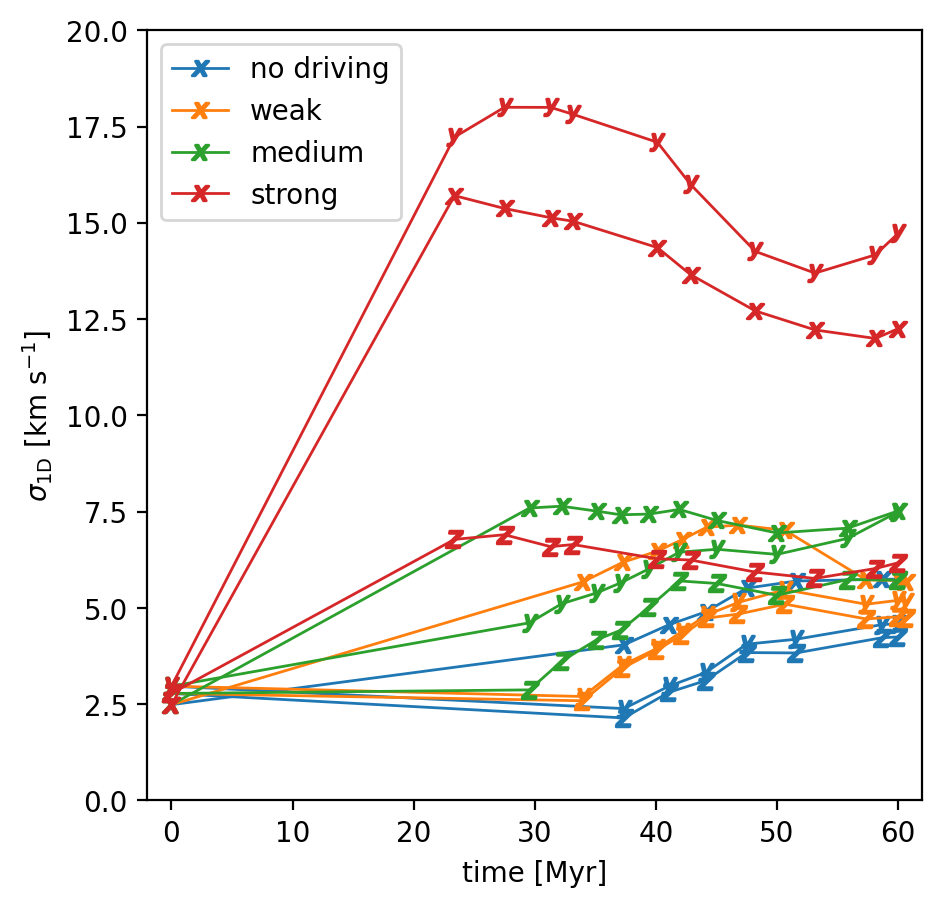}
    \caption{The mass-weighted velocity dispersion within the disk (midplane $\pm$ 200 pc) throughout the simulations}
    \label{fig:sigma}
\end{figure}

\begin{figure}
    \centering
    \includegraphics[width=\columnwidth]{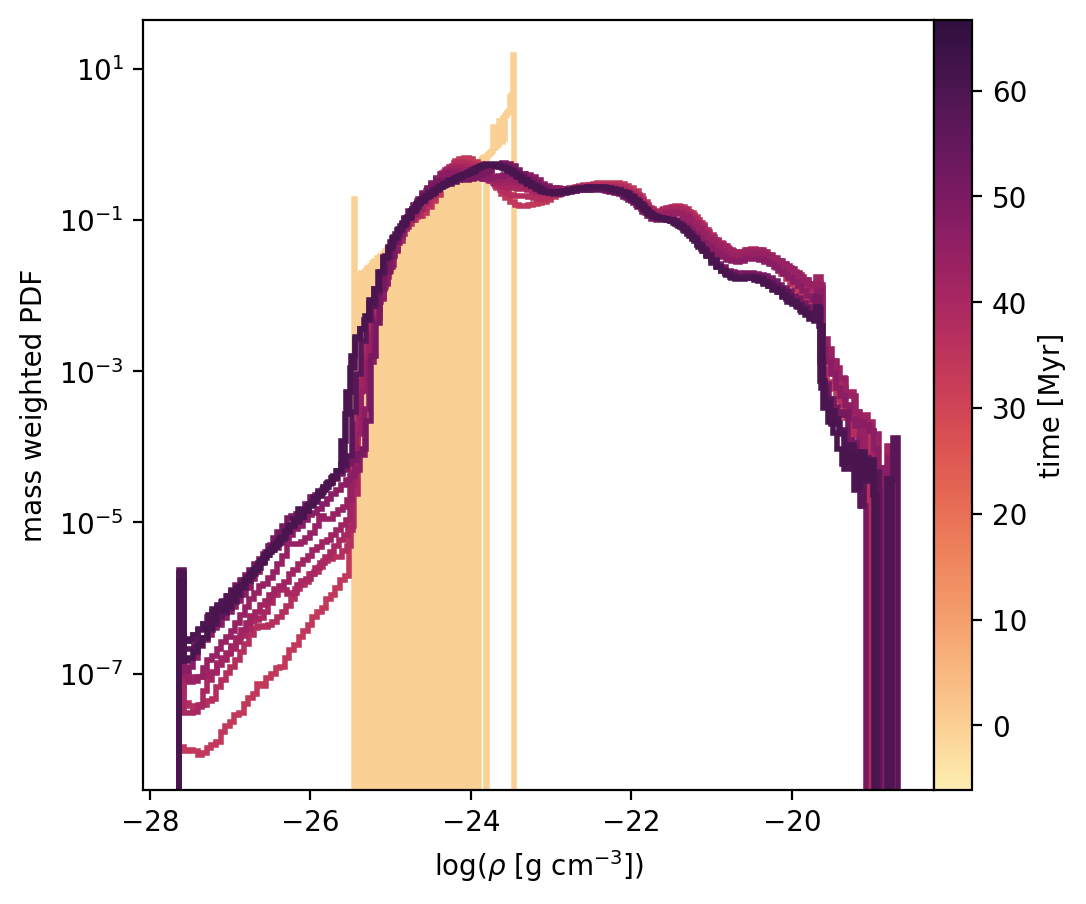}
    \caption{The evolution of the density PDF in the simulation with weak driving.}
    \label{fig:pdf_evo}
\end{figure}

Before analyzing the final snapshots of our simulations, we investigated how our simulations evolve over time.
This is to make sure that the simulations have passed the initialisation phase, the turbulence is well developed and we have reached statistical stationary.

Figure~\ref{fig:sigma} shows the evolution of the mass-weighted velocity dispersion within the disk ($\pm$ 200 pc from the mid-plane to be exact).
Apart from in the strong driving case, the velocity dispersion converges to a roughly constant value.

Figure~\ref{fig:pdf_evo} shows the evolution of the mass weighted density PDF for the simulation with weak driving. This converges to a roughly constant form by the end of the simulation. The behaviour of the density PDF in the other simulations is very similar. Together with the lack of significant variation in the velocity dispersion, this reassures us that our simulations are well-evolved and we can proceed to analyse the final snapshots.


\bsp	
\label{lastpage}
\end{document}